\documentclass[a4paper]{emulateapj}

\usepackage[english]{babel}
\usepackage[utf8x]{inputenc}
\usepackage[T1]{fontenc}
\usepackage{color}
\usepackage[dvipsnames]{xcolor}

\usepackage{graphicx}

\newcommand{\msun}{\ensuremath{\,M_\odot}}

\newcommand{\zsun}{\ensuremath{\,Z_\odot}}

\newcommand{\yr}{\ensuremath{\,\mathrm{yr}}}

\newcommand{\myr}{\ensuremath{\,\mathrm{Myr}}}
\newcommand{\gyr}{\ensuremath{\,\mathrm{Gyr}}}

\newcommand{\ergs}{\ensuremath{\,\mathrm{erg}\,\mathrm{s}^{-1}}}
\newcommand{\msy}{\ensuremath{\msun\mathrm{\; yr}^{-1}}}

\newcommand{\startrack}{{\tt StarTrack}}

\newcommand{\new}[1]{{#1}} 

\begin{document}

\title{The observed vs total population of ULXs}
%
%
%
%

\author{
		Grzegorz Wiktorowicz\altaffilmark{1,2}\thanks{E-mail: gwiktoro@astrouw.edu.pl},
        Jean-Pierre Lasota\altaffilmark{3,4},
        Matthew Middleton\altaffilmark{5},
        Krzysztof Belczynski\altaffilmark{3}
}

 \affil{  
    $^{1}$ National Astronomical Observatories, Chinese Academy of Sciences, Beijing 100101, China\\
    $^{2}$ School of Astronomy \& Space Science, University of the Chinese Academy of Sciences, Beijing 100012, China\\
    $^3$ Nicolaus Copernicus Astronomical Center, Polish Academy of Sciences, Bartycka 18, 00-716 Warsaw, Poland\\
    $^4$ Institut d’Astrophysique de Paris, CNRS et Sorbonne Université, UMR 7095, 98bis Bd Arago, 75014 Paris, France\\
    $^5$ Department of Physics and Astronomy, University of Southampton, Highfield, Southampton SO17 1BJ, UK
}
            
\begin{abstract}
   \centering
	We have analyzed how anisotropic emission of radiation affects the observed sample of ultraluminous X-ray sources (ULXs) by performing simulations of the evolution of stellar populations, employing recent developments in stellar and binary physics, and by utilizing a geometrical beaming model motivated by theory and observation.  Whilst ULXs harboring black hole accretors (BH ULXs) are typically emitting isotropically, the majority of ULXs with neutron star accretors (NS ULXs) are found to be beamed. 
 These findings confirm previous assertions that a significant fraction of ULXs are hidden from view due to a substantial misalignment of the emission beam and the line-of-sight. {\bf We find the {\it total} number of NS ULXs in regions with constant star formation, solar metallicity, and ages above $\sim1\gyr$ to be higher than the BH ULXs, although {\it observationally} both populations are comparable. For lower metallicities BH ULX dominate both the total and observed ULX populations. As far as burst star-formation is concerned, young ULX populations are dominated by BH ULXs, but this changes as the population ages and, post star-formation, NS ULXs dominate both the observed and total ULX populations}. We also compare our simulation output to a previous analytical prediction for the relative ratio of BH to NS ULXs in idealized flux-limited observations and find broad agreement for all but the lowest metallicities. In so doing we find that in such surveys the observed ULX population should be heavily dominated by black-hole systems rather than by systems containing neutron stars.

\end{abstract}

\section{Introduction}

Geometrical beaming occurs in an accreting system when radiation preferentially  escapes along a beam with an opening solid angle $<$ 4$\pi$ steradians. As a result, an observer located in the cone of emission will infer a higher luminosity by assuming isotropic emission, than the real total integrated luminosity of the system. There are observational and theoretical arguments for the presence of beaming in systems with very high accretion rates onto a compact object. For example, despite its very high accretion rate \citep[$\dot{M}\approx10^{-4}$ $M_{\odot}$ year$^{-1}$][]{Fabrika04}, the binary system SS 433 is faint in the X-rays ($L_{\rm X}\approx10^{36}\ergs$) with recent evidence \citep{Middleton1810} indicating that most of the radiation escapes at high inclinations to the line of sight \citep[similar to the jets,][]{Begelman0607,Medvedev1002}.  
%
%
%
%

From a theoretical point of view, \new{ basic physics} and detailed numerical calculations indicate that at super-Eddington accretion rates, the disk ceases to be geometrically thin around the spherisation radius \new{\citep[$R_{\rm sph}$, e.g.][]{Shakura73,Ohsuga0906,Ohsuga1107,Sadowski1403}} \new{ where radiation pressure dominates} and the Eddington limit is reached locally. The location of $R_{\rm sph}$ is expected to be linearly proportional to the accretion rate,  $R_{\rm sph}\sim \dot{m}$, where $\dot{m}$ is the mass accretion rate in Eddington units,  $\dot{m}_{\rm Edd} = L_{\rm Edd}/\eta c^{2}$, with a radiative efficiency, $\eta \approx$ 0.1,  and an Eddington luminosity ($L_{\rm Edd}$) for an accretor mass ($M_{acc}$) and hydrogen abundance in the accretion flow ($X$):

%
%

\begin{equation}\label{eq:eddlim}
L_{\rm Edd}=2.6\times10^{38}\frac{1}{1+X}\frac{M_{\rm acc}}{\msun} ~~ \left[ \frac{\rm erg}{\rm s}\right].
\end{equation}

Due to the large aspect ratio of the disc and ease with which material is lost in a wind (needed to keep the accretion rate at the Eddington value for smaller radii in the absence of advection), emission from the inner-most regions (where the most energetic photons are formed) is trapped in a conical, optically thick structure. 

%
%
%
%
%
%
%
%

Ultraluminous X-ray sources (ULXs) are defined as point-like, off-nuclear sources with -- isotropically assumed -- observed X-ray luminosities above $L_{\rm ULX}=10^{39}\ergs$ \citep[for a recent review see][]{Kaaret1708} and are particularly important in the context of super-Eddington accretion. One interpretation involves sub-Eddington accretion onto intermediate-mass BHs \citep{Colbert9907}. However, \citet{King0105} showed that globular clusters on average cannot produce the necessary number of IMBHs to explain all ULXs and, instead argue that only the presence of beaming in ULXs avoids serious formation difficulties. Observationally, evolution in the X-ray spectra of ULXs, coupled with the short timescale variability would also argue for geometrical beaming in a super-critical flow \citep{Middleton1503,Middleton1610}.
%
%

\new{The population synthesis of ULXs has been performed in several studies. \citet{Rappaport0501} showed that population of ULXs in spiral galaxies can be explained by short high mass transfer phases in BH+MS binaries. However, they neglected pre-supernova evolution in their calculations and didn't take into account other types of binaries, e.g. containing NS accretors, or evolved donors. Their study was expanded in \citet{Madhusudhan0812} where they additionally predicted the observational properties of the ULX population. Meanwhile, intermediate mass BHs were proposed as potential (if hypothetical) accretors in ULXs for which a violation of the Eddington limit is not necessary \citep{Colbert9907,Madhusudhan0604}. More recently, \citet{Linden1012} utilized the \startrack\ population synthesis code to show that the bulk of ULXs can be explained as a high-luminosity tail of  high-mass X-ray binaries. Since the recent discovery of NS in ULXs \citep{Bachetti1410},  NS accretors were included in studies of ULXs. For these objects, the Eddington limit is apparently surpassed more than several times \citep[e.g.][]{Fragos1503,Shao1504,Wiktorowicz1509}. Similarly, the detection of double compact object mergers \citep{Abbott1602} triggered an investigation of potential connections between double compact objects and ULXs \citep[e.g.][]{Marchant1708,Finke1712,Klencki1811}. Massive stars (mainly red super-giants) were detected in optical and infrared bands as potential donors in a few ULXs \citep[e.g.][]{Liu0402,Kaaret0407,Heida1408,Heida1606}, whereas \citet{Wiktorowicz1709} predicted main sequence donors (typically $5.9$--$11\msun$ for BH accretors and $0.9$--$1.5\msun$ for NS accretors) for the majority of ULXs.}

The discovery of pulsing ULXs \citep[PULX;][]{Bachetti1410} has called into question the role of beaming versus strong dipole magnetic fields \citep[e.g.][]{Mushtukov1502}, but observations of cyclotron resonance lines \new{\citep[\citeauthor{Brightman1804} \citeyear{Brightman1804}, \citeauthor{Walton1804} \citeyear{Walton1804}, but see][]{Koliopanos1901}} would broadly support "normal'',  pulsar-like dipole field strengths $10^{11}  \lesssim B \lesssim 10^{13}$G. The combination of such field strengths and super-Eddington accretion rates, allows for a consistent -- if not complete -- description of the observed properties of PULXs \citep[Middleton et al. in prep.]{King1605,King1702,King1905}. In particular,  \citet{King1605} found that neutron star ULXs (NSULXs) are likely to have higher apparent luminosities than black hole ULXs (BHULXs) for a given mass transfer rate, as their increased beaming outweighs their lower Eddington luminosities. For example, using methods provided in Sec.~\ref{sec:methods}, for a typical NS and BH mass \citep[$1.4\msun$, and $7\msun$, respectively e.g.][]{Ozel1012} and a mass transfer rate $\dot{M}=10^{-5}\msy$, the real,  total integrated luminosity ($L_{\rm X}$) is higher for the BH ($\sim5.3\times10^{39}\ergs$) than for a NS  ($\sim1.4\times10^{39}\ergs$). However, the apparent luminosity for an observer located in the emission cone is higher for the NS ($\sim4.3\times10^{41}\ergs$) than for the BH  ($\sim2.0\times10^{41}\ergs$).
\new{We note, that some GRMHD simulations show the opposite result. For example, \citet[][]{Abarca1809} performed a simulation of super-Eddington accretion ($\dot{M}=200 L_\mathrm{Edd}/c^2$) onto a non-magnetized non-rotating neutron star ($M_\mathrm{NS}=1.4\msun$) and, contrary to the BH case, found no significant beaming. In this model the luminosity is about $L_\mathrm{X}\approx10^{38}\ergs$ and pulsations would not be visible, and so is not directly related to PULXs.}
%
%
%
%
%
%
%
%
Although \new{magnetar-strength} magnetic fields may also be responsible for the emergence of super-Eddington levels of radiation \citep[e.g.,][]{Basko7605,Mushtukov1512} \new{and strongly influence the mass flow in the accretion disk \citep[e.g.][]{Parfrey1712}}, in this paper we focus on geometrical beaming only. 
\new{We are encouraged to do so by the fact that {\sl all} known magnetars are single \citep{Olausen0514} and although they are supposed to be formed in binaries, their birth leads to the binary orbit disruption \citep[see e.g.,][]{Clark0514}. The rare survivors are unlikely to be present in PULXs \citep{Popov0116}.}

For a given funnel opening angle $\theta$, the  probability of an observer being located in its cone of emission is given by:
%
%
%
%
\begin{equation}\label{eq:pobs}
P_{\rm obs}(\theta)=1-\cos\theta/2,
\end{equation}
Therefore, the stronger the beaming (lower $\theta$), the lower the fraction of {\it observed} systems in the total population.  On the other hand, beamed sources may be visible from much larger distances due to higher apparent luminosities \citep[see][Sec.~\ref{sec:fluxLimitedSurvey}]{Middleton1708}. 

In this paper, we analyze the impact of beaming on the relation between the observed and total sample of ULXs. Our calculations are based on results presented in \citet{Wiktorowicz1709} where beaming was already included, but not analyzed in detail.
Our main motivations are the recent discoveries of NSs in ULXs \citep{Bachetti1410,Israel1609,Furst1611,Carpano1805} and observational hints that many non-pulsing ULXs may host NSs \new{\citep[e.g.][]{Pintore1702,Mushtukov1705,Koliopanos1712,Walton1804}} as predicted by previous works, e.g. \citet{King0105} and \citet{King1605}.
%
%
%
%

\section{Methods}
\label{sec:methods}

In \citet{Wiktorowicz1709}, ULX populations in different environments were analyzed, however, that work focused only on {\it observed} ULXs, i.e. those which are predicted to be visible from the Earth. The total population of ULXs \citep[including so called "misaligned", or "hidden" sources such as SS433: \citeauthor{Middleton1810}~\citeyear{Middleton1810}, or MQ1 in M83:][]{Soria1403}  was not analyzed. 

We utilized the {\tt Startrack} population synthesis code \citep{Belczynski0206,Belczynski0801} with further updates \citep[see][and references therein]{Wiktorowicz1709}. For the initial primary masses  we used the Kroupa IMF with ${\rm P}(M_{\rm ZAMS})\propto M_{\rm ZAMS}^\Gamma$\citep{Kroupa0312} across a range $5$--$150\msun$. The power-law index, $\Gamma=-1.3$ for stars with $M_{\rm ZAMS} \leq 0.5\msun$, $\Gamma=-2.2\msun$ for stars with $0.5<M_{\rm ZAMS}\leq1$, and $\Gamma=-2.3$ for stars heavier than $1\msun$ on ZAMS. The distribution of mass ratios ($q=M_2/M_1$, where $M_{1/2}$ is the primary/secondary mass) was assumed to be uniform between $q_{\rm min}=0.08\msun/M_1$ and $1$.  The initial distribution of orbital periods ($P$) and eccentricities ($e$) are  ${\rm P}(\log P)\sim(\log P)^{-0.55}$ and ${\rm P}(e)\sim e^{-0.42}$ \citep{Sana1207}, which is the main difference in comparison to \citet{Wiktorowicz1709}, but strong differences in the resulting binary populations are not expected \citep[]{deMink1511,Klencki1811}. 
%
%
%
%
%
%
%
%
%
%
%
%

Every binary formed is evolved over $10\gyr$ in isolation, i.e. no dynamical interactions with third bodies are taken into account, with special attention paid to interactions such as common envelope \citep[CE;][]{Ivanova1302} and mass transfer \citep[MT; for details see][]{Belczynski0801}. To estimate the final compact object mass after a supernova, we use the ``rapid'' supernova formation mechanism \citep{Fryer1204,Belczynski1209}. For both NSs and BHs, we draw natal kicks from a Maxwellian distribution with $\sigma=265$ km/s \citep{Hobbs0507}, but scaled proportionally to the fraction of ejected mass which falls back onto the compact object. The kick velocity applied to a newly formed compact object ($v_{\rm kick,fin}$) is obtained from $v_{\rm kick,fin}=v_{\rm kick} (1-f_{\rm fb})$, where $v_{\rm kick}$ is the kick velocity that was drawn from a Maxwellian distribution with $\sigma =265$ km/s, and  $f_{\rm fb}$ is the fraction of mass that was ejected in the SNa explosion that is
accreted back onto the compact object. We assumed that  BHs forming via direct collapse obtain no natal kick.

%
%
%
%
%
%
%
%
We focus exclusively on sources undergoing Roche lobe overflow (RLOF) mass transfer (MT), during which, the X-ray luminosity is assumed to be \citep{Shakura73,Poutanen0705}:
\begin{equation}\label{eq:Lx}
L_{\rm X}=\left\{\begin{array}{ll}
L_{\rm Edd}(1+\ln \dot m_{\rm tr})&\dot m_{\rm tr}>1 \\
L_{\rm Edd}\dot m_{\rm tr} &\dot m_{\rm tr}\leq1
\end{array}\right.,
\end{equation}
where $\dot m_{\rm tr}=\dot{M}_{\rm tr}/\dot{M}_{\rm Edd}$ is the MT rate in Eddington units and $\dot{M}_{\rm tr}$ is the mass transfer rate. The 
%
%
The {\it apparent} (spherical) luminosity is:
\begin{equation}
L_{\rm app} = L_X/b,
\label{eq:Lapp}
\end{equation}
%
%
%
%
%
%
%
%
%
%
%
%
%
%
%
%

\noindent where the beaming factor is defined as $b\stackrel{def}{=}\Omega/4\pi=P_{\rm obs}(\theta)$ and $\Omega$ is the combined solid angle of both beams. \citet{King0902} showed that the observed relation of soft X-ray excess ($L_{\rm soft}$) and disk temperature ($T_{\rm disk}$) in ULXs, $L_{\rm soft}\propto T_{\rm disk}^{-4}$, implies:
\begin{equation}
	b\sim\frac{73}{\dot{m}_{\rm tr}^{2}} \,\,\,\mathrm{for}\,\, \dot{m}_{\rm tr} > 8.5,
    \label{eq:beamingKing}
\end{equation}
whereas for $\dot{m} < 8.5$, emergent radiation is essentially unbeamed.  In the following, the beaming is included as:
\begin{equation}\label{eq:beaming}
b=\left\{\begin{array}{ll}
1 & \dot{m}_{\rm tr} \leq 8.5,\\
\frac{73}{\dot{m}_{\rm tr}^2} & 8.5<\dot{m}_{\rm tr},
\end{array}\right.
\end{equation}
%
%
%
%
\noindent which provides monotonicity and continuity. These formulae have been successfully used to describe various classes of ULXs and hyperluminous X-ray sources \citep[HLXs; see e.g.,][]{King1410,King1605,King1702,Lasota1503}. 
%
%
%
%

\new{The prescription Eq. (\ref{eq:beamingKing}) for the beaming factor is supported both by observations and theory \citep{King0105,King0902}.  \citet{Miller1013} analysed a sample of ULXs finding them to broadly adhere to a $L_{\rm soft} \propto T^{4}$ relation. Whilst this may be due to freezing the absorption column -- which, in at least one object, is observed to change, possibly with precession phase \citep{Middleton1512} -- it may also indicate a changing fraction of energy lost via winds or very massive black holes very close to the Eddington limit.}
In \citet{Wiktorowicz1709} we assumed an {\it ad hoc} beaming saturation in order to avoid exceedingly small values of $b$ (and correspondingly large values of $L_{\rm app}$). In the context of the present paper this is not necessary because, as we have confirmed, extremely beamed sources are not only hard to observe, but also extremely rare and short-lived.

\new{In \citet{Wiktorowicz1709} we have tested different beaming models applicable to population synthesis: a model with no beaming at all, a model with constant beaming ($b=0.1$) for all super-Eddington sources, and a model based on the relation of the photosphere height to disk radius $H/R=1.6/1+\frac{r}{\dot{m}}$ obtained with the GRMHD code {\tt KORAL} \citep{Lasota1603}. The resulting numbers of ULXs, and ratio of BH to NS ULXs are highly similar (differing by a factor of $\lesssim2-3$), except in the case of the least physical model with constant beaming, where the differences were more significant.

The results from other detailed GRMHD simulations, which are not applicable to population synthesis due to scarcity of tested configurations, frequently show results which diverge from the prescriptions of \citet{King0105} \citep[e.g.][]{Sadowski1502}. In these simulations a highly colimated outflow never forms and the beam remains wide even for very high mass transfer rates \citep{Lasota1603}. Similar problem we see when analysing the simulation of super-Eddington accretion on super-massive BHs \citep[e.g][]{Dai1806}, which are relevant here due to scale-free behaviour of the most important relations (except for the radiation to gas pressure ratio). Nevertheless, these codes give results comparable to these obtain with the {\tt KORAL} code, so, as stated above, we predict a small effect on our results and especially on our conclusions. The only situation where the results concerning ULX are significantly different from detailed codes and prescription of \citet{King0902} are the most luminous ULXs \citep[so called extreme ULXs;][]{Wiktorowicz1509}, but as we have already said, these extreme ULX are unimportant for our general conclusions.}

\new{It should be stressed, however, that the \citet{King0902} scenario, uses the \citet{Shakura73} model in which an accretion flow
fed at a super-Eddington {\sl mass-transfer} rate consists of an external, geometrically--thin disk down to
the spherization radius, below which the mass {\it accretion} is effectively only Eddington,
the excess being ejected (advection of energy is negligible) in a quasi--spherical outflow whose collimating structure might be the cause
of the anisotropic luminosity. In the case of neutron star accretors, this picture is strongly supported by the fact that the low--mass X--ray binary Cygnus X--2, has survived
being fed $\sim 3\msun$ from its (initially more massive) companion star at very
super--Eddington rates ($\sim 10^{-5}\msun\, {\rm yr}^{-1}$), but has evidently gained no more 
than $\sim {\rm few}\times 0.2\msun$ \citep{King9910}, evidently 
ejecting all the surplus. On the other hand, none of the GRMHD codes simulates the external thin disk but
uses a torus outside the inner $\sim 50$ gravitational radii so their results are not directly comparable with
the \citet{Shakura73} model, since e.g., the definitions of the spherization radii are not comparable.
To make things even more confusing, the luminosity formula $L=L_{\rm Edd}(1+\ln \dot m_{\rm tr})$ is the same
for the \citet{Shakura73} and the advection dominated models \citep[][]{Lasota1505,Poutanen0705} but in the first
case $\dot m(R) = \dot m_{\rm tr}$, whereas in the second $\dot m(R) \propto R$.
}

\section{Results I - Volume-limited surveys}

The results presented in this section are for a uniform population of initial binaries (same metallicity of all stars: either solar  $Z=\zsun=0.02$, $10\%\zsun=0.002$, or $1\%\zsun=0.0002$) and assuming a simple model of star-formation (SF): a constant star formation rate (SFR), with duration of $10\gyr$, or burst-like, with a duration of $100\myr$. Although both models form a total stellar mass of $6\times10^{10}\msun$, which corresponds to the stellar mass of the Milky Way \citep[including bulge and disk;][]{Licquia1506}, they cannot be directly compared to the Milky Way, or any other complex stellar system which have various episodes of star formation and not uniform chemical composition. Nevertheless, estimates for more realistic systems may be obtained through use of our presented results\footnote{Data files will be soon available at \url{https://universeathome.pl/universe/ulx.php}. In case of questions concerning the usage, please contact the first author.}.

We focus on a volume-limited case in which we constrain the volume within which we observe ULXs.  This then applies to volume-limited catalogues  \citep[e.g.][]{Walton1109,Swartz1111}, or galaxy-focused observations \citep[e.g.][]{Wolter1808}.  We note that the same volume of space may contain different amounts of stellar mass, (or, equivalently, the same stellar mass may occupy different volumes). In the following, we use the volume that contains the stellar mass of the Milky Way. For comparison with observations containing more/less stellar mass, the results should be scaled up/down accordingly, as the number of ULXs is directly proportional to stellar mass.

\begin{figure}[t]
    \centering
    {continuous star formation}\\
    \includegraphics[width=1.0\columnwidth]{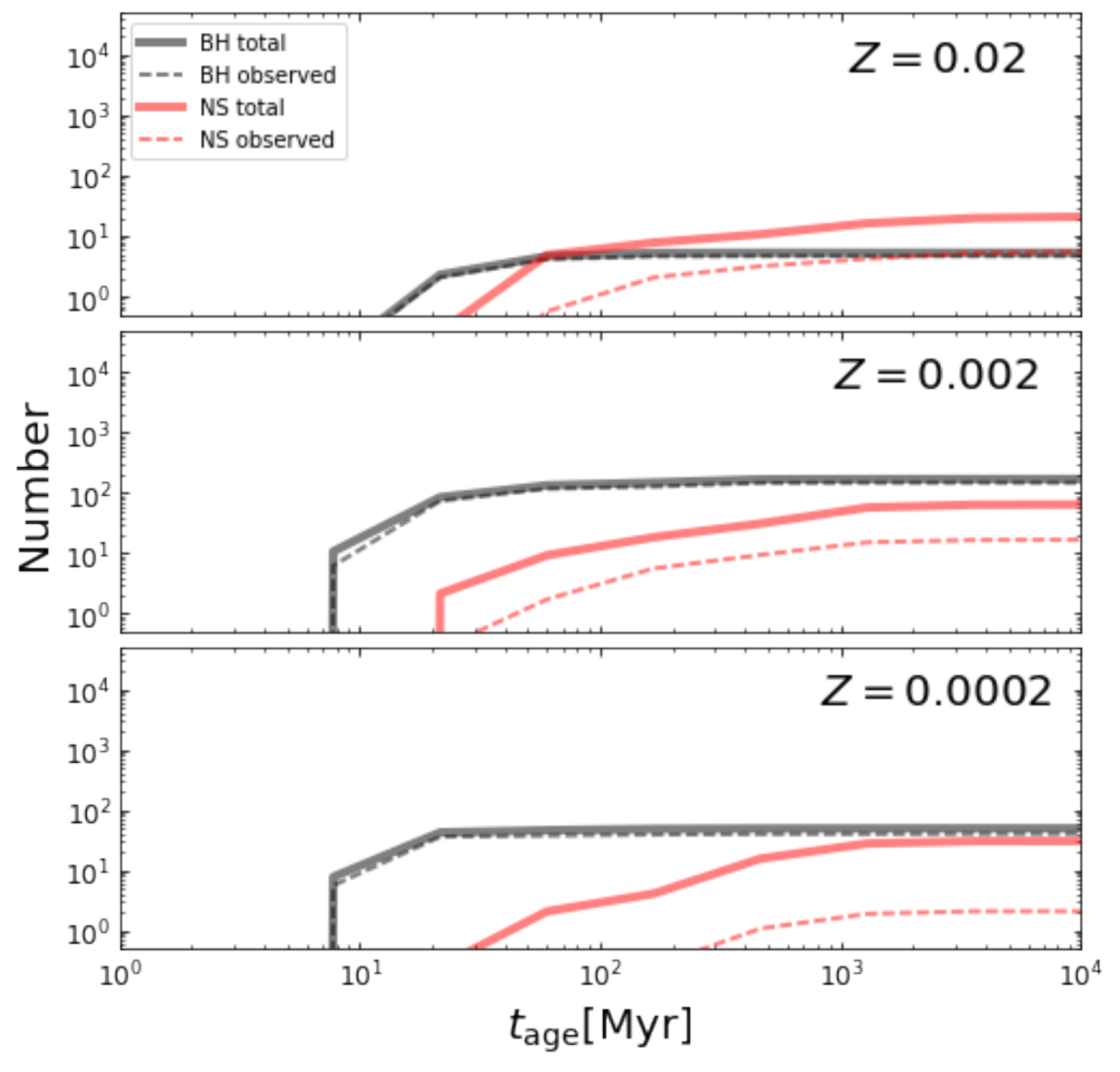}\\
    \vspace{0.5cm}
    {burst star formation}\\
    \includegraphics[width=1.0\columnwidth]{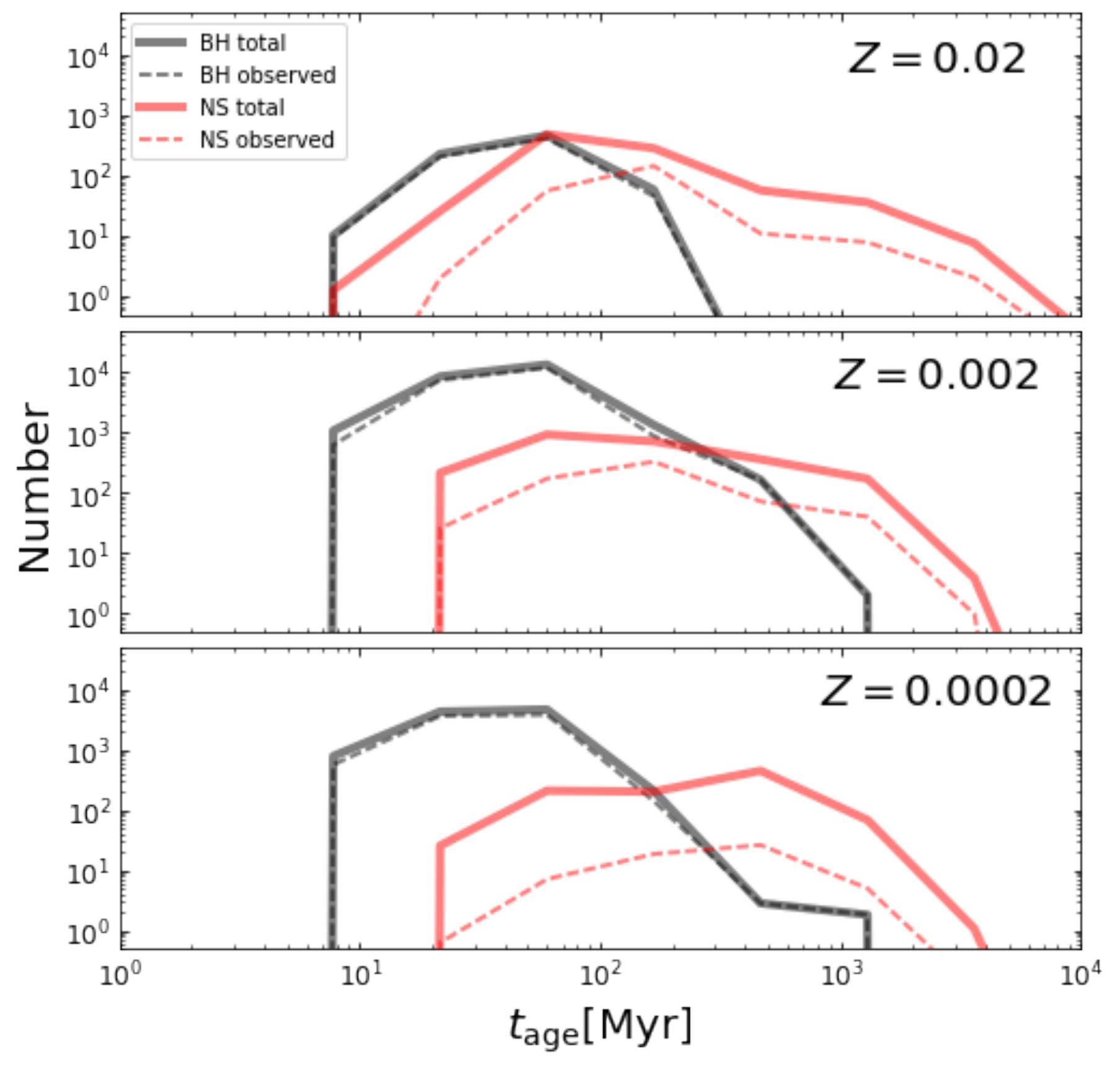}
    \caption{Number of ULXs as a function of time since the beginning of star formation. The three upper panels present the results for constant star formation, whereas the three lower ones are for a star formation burst which lasted $100\myr$. The total stellar mass formed is the same for all plots and equals $6\times10^{10}\msun$, which is approximately the total stellar mass of the Milky Way galaxy \citep{Licquia1506}. Three metallicities were considered (solar  $Z=\zsun=0.02$, $10\%\zsun=0.002$, and $1\%\zsun=0.0002$). On each panel, four lines present the total number of ULXs with BH, or NS accretors (gray solid and dashed line, respectively) and observed ULXs (red lines), i.e. ULXs whose beam intercepts the Earth (see Eq.~\ref{eq:pobs}). Although for BH ULXs the difference between visible and total sample is small, for NS ULXs the total sample is typically $5$--$15$ times larger than the observed one.}
    \label{fig:num}
\end{figure}

Our results are presented in Fig.~\ref{fig:num}. This is an updated and expanded version of Fig.~2 from \citet{Wiktorowicz1709}, where only the observed ULXs were presented. Here we also show the total number of ULXs (i.e. the sum of both observable and hidden sources; based on Eq.~\ref{eq:pobs}).


\begin{figure}[t]
    \centering
    \includegraphics[width=1.0\columnwidth]{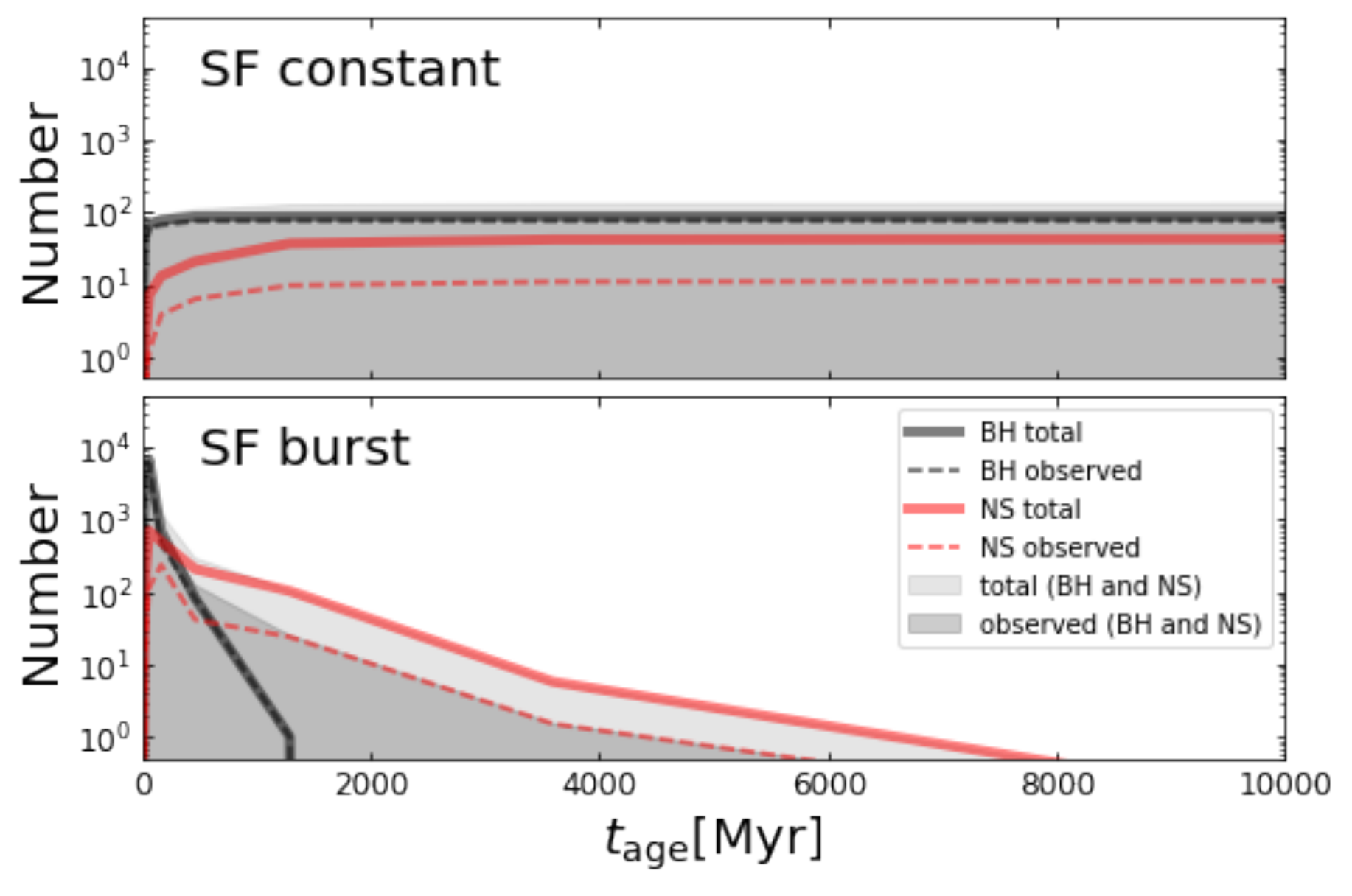}
    \caption{The toy model of the Milky Way galaxy constructed as a mix of 50\% solar metallicity stars ($Z=0.02$) and 50\% sub-solar metalicity stars ($Z=0.002$). Although the SFH of the Galaxy is much more complicated, here we present two simple cases: constant SF and burst SF (similarly to Fig.~\ref{fig:num}) and discuss them in the text.}
    \label{fig:toy}
\end{figure}
%
%
%
%

In order to demonstrate how our results can be applied for practical purposes, we have constructed a simple ("toy") model of the Galaxy. We assumed that $50\%$ of its stars have formed with solar metallicity ($Z=0.02$) and $50\%$ with a lower one ($Z=0.002$). If we impose a constant SFR (Fig.~\ref{fig:toy}; upper panel), the resulting synthetic population ($\sim100$ observed ULXs) is strongly inconsistent with the actual one (no {\it observed} ULXs). However, measurements of the star formation history (SFH) in the Milky Way suggest that the SFR may have been significantly higher in the past \citep[$\gtrsim$ {\it a few} Gyr ago, e.g.][]{SilvaAguirre1804}. The SFH is, therefore, more burst-like. For such a case (Fig.~\ref{fig:toy}; bottom panel) we predict that only $2$ Gyr after the burst,  more than $80\%$ of ULXs are hidden from our view, for which SS433 may be a representative \citep{Middleton1810}.
%
%
%
%
%
%
%
%
%
%

Nevertheless, observations also suggest recent, small but significant SF in the Galaxy \citep[$\sim1\msun/\yr$ for $\sim1\gyr$, e.g.][The burst which might have happended $\sim2\gyr$, or ealier bursts have little effect on the number of ULXs; see Fig.~\ref{fig:num}; lower panels]{Maciel1208}, which, according to the "toy" model, should produce $\sim20$ (or $\sim3$ if the SF occurs only in metal-rich environments; Fig.~\ref{fig:num}, upper-most panel) observable ULXs at the current time. In contrast, in the Milky Way we observe only sources which become ULXs during their outburst's peaks when their emission goes above $10^{39}\ergs$, such as the BeX binary Swift J0243.6+6124 \citep[$L_{\rm X, peak}\approx5\times10^{39}\ergs$;][]{Tsygankov1702} or some low-mass-X-ray-binary transients \citep[LMXBTs;][]{Tetarenko1602}. These "transient ULXs" are not a part of our results because BeX binaries do not belong to our sample and the mass-transfer rates of the LMXBTs are well below the Eddington value.

A better agreement with observations is obtained when we do a simple scaling of our results to the observed stellar mass of galaxies within $14.5$ Mpc ($M_{\rm tot}\approx3.5\times10^{12}\msun$) where $107$ ULXs were found \citep{Swartz1111}. Assuming that in this volume the recent ($\sim1\gyr$) SFR was small $\sim1\msun/\yr$  and chemical composition in this volume is similar to these used in the "toy" model, we obtain a prediction of $175$ ULXs, which is a less then a factor of two difference. 

Possible sources of discordance include both theory and observation. It was shown in, e.g.,  \citet{Wiktorowicz1709} that the predicted number of ULXs may vary depending on the accretion model used. However, the ratio of NS to BH ULXs is only slightly affected, which agrees with our general conclusions. Some evolutionary phases which are important for the formation of ULXs (and X-ray binaries in general) like the CE  are not well understood and we do not have good models for them. Nevertheless, in \citet{Wiktorowicz1409}, we showed that different CE models, which give significantly different CE outcomes, result in very similar predictions for X-ray binary (including ULX) populations, although populations of progenitors may differ significantly. Therefore, other evolutionary models may improve the fit to the simple model of the Milky Way presented above, but will not change our general conclusions. We also note that estimates of the observational parameters, which are necessary for population synthesis studies, like the total stellar mass and SFH are not very precise. For example, the recent estimates of the Milky Way's stellar mass vary by a factor of $\sim2$ \citep[compare, e.g.,][]{Bovy1312,Licquia1506,Xiang1808}.

\subsection{Ratio of observed to total sample}

%
%


The initial NS mass in ULXs according to our simulations is typically around $1.3\msun$. After  formation, the NS's mass increases due to accretion, and in the ULX population is typically around $1.4\msun$. If the ULX phase occurs early after ZAMS ($t_{\rm age}\lesssim500\myr$), donors are typically MS stars, and HG/RG (HG = Hertzsprung gap, RG = red giant) if it occurs later ($t_{\rm age}\gtrsim1\gyr$). Typically, the donor mass is $M_{\rm don}\lesssim2\msun$.  

Within our simulation, we find that the majority of NS ULXs are beamed sources (Figs.~\ref{fig:bfrac} and~\ref{fig:bfrac_obs}) which results in a low average ratio of the observed to the total number of NS ULXs ($1/15$--$1/5$, depending on the metallicity, Fig.~\ref{fig:num}). The beaming results from the fact that a NS typically requires a strongly supercritical accretion rate to appear as a ULX because sub-critically accreting NSs will have Eddington-limited apparent luminosities  ($L_{\rm X}\leq L_{\rm Edd,NS}\lesssim5\times 10^{38}\ergs$), well below the empirically defined ULX luminosity of $L_{\rm ULX}=10^{39}\ergs$. Indeed, a typical NS observed in a binary system ($M_{\rm NS}\approx1.4\msun$) requires  $\dot{m}>10$ to reach $L_{\rm app}>L_{\rm ULX}$ (see Eq.~\ref{eq:Lapp}), so the required beaming factor is always lower than $b\lesssim0.7$.  This means that the probability of observing a typical NS ULX is always lower than $\sim70\%$. Our results indicate that the average probability is, actually, only between $\sim7$--$20\%$. 


Exceptions to the above do occur; we note that some NS ULXs may emit isotropically, especially when the metallicity of the environment is $Z\gtrsim10\%\zsun$.  In these cases, a NS may undergo a long phase of MT during which its mass increases to $\sim1.8\msun$ and the donor loses its hydrogen envelope. In such systems, the donor is typically a low-mass ($\sim0.1\msun$) hybrid WD with a C-O-He rich core and a He rich envelope, in a very close orbit with the NS \citep[orbital period $P<1$ h; see][]{Belczynski0403}. Noting that the Eddington limit is higher for helium rich donors (Eq.~\ref{eq:eddlim}) implies that the isotropic emission of such a system may surpass $L_{\rm ULX}$ when $\dot{m}\gtrsim3$, which is significantly lower than required for a NS with a typical mass of $1.4\msun$ accreting from a hydrogen-rich donor. These systems are good candidates for the brightest ultra-compact X-ray binaries \citep[cf.][]{King1105}, however, their fraction among NS ULXs is only rarely expected to be higher than $4\%$ (see Fig.~\ref{fig:bfrac})

Unlike the condition for NSs,  as the defining ULX luminosity  ($L_{\rm ULX}=10^{39}\ergs$) is the Eddington luminosity for a $\sim7\msun$ BH --  a typical mass of a stellar-mass BH in the Milky Way galaxy \citep[e.g.][]{Ozel1012} -- a large fraction of BHs can become ULXs {\it without} the need for highly super-critical MT rates. When we take into account the lack of beaming in our models up to $\dot{m}<8.5$, binaries with all stellar-mass BH may obtain ULX luminosities without additional amplification. According to the adopted beaming model (Eq.~\ref{eq:beaming}), these objects will emit isotropically, i.e. $P_{\rm obs}=1$. 
%
%
%
%
%
%
%
%
%
%
%
%
However, a fraction of the BH ULX population {\it may} be beamed due to highly super-critical MT rates and naturally results in high apparent luminosities ($\gtrsim10^{40}\ergs$; see Fig.~\ref{fig:num}). 
Typically, the beamed fraction is  $\lesssim10\%$ of the total population (e.g., Fig.~\ref{fig:bfrac}, $Z=0.02$), but it may exceed $50\%$ in extreme cases (e.g., Fig.~\ref{fig:bfrac}: $Z=0.002$, shortly after star formation ceases) and can reach nearly $100\%$ for extremely young populations (Fig.~\ref{fig:bfrac}: $t_{\rm age}<5\myr$).
%
%
%
%
BH ULXs most commonly appear in low metallicity environments ($Z\ll\zsun$) and very young stellar populations ($t_{\rm age}\lesssim6\myr$), and the majority of these systems will still have MT rates below $\dot{m}\approx 8.5$, because they are easier to obtain as a result of binary evolution. These "low-luminosity" BH ULXs \citep[see][]{Middleton1203,Middleton1301} outweigh the extreme (beamed) BH ULXs and, as a consequence, the ratio of the observed to the total sample of BH ULXs is typically only $\sim0.8$. The average probability of observing a BH ULX from the Earth is therefore  $\sim80\%$,  far higher than the case for NS ULXs discussed above. 

A current, important question in the field is "What is the ratio of observed NS to BH ULXs?" and our simulations allow us to consider this under our model assumptions. For star-burst systems (Fig.~\ref{fig:num}, bottom panels), during star formation, BH ULXs dominate the total and observed ULX population, while NS ULXs dominate these populations once star formation has ceased. For prolonged/continuous star formation (Fig.~\ref{fig:num}, top panels), BH ULXs always dominate the total and observed ULX populations for low metallicities ($Z=0.002$ and $Z=0.0002$ models). For high metallicity ($Z=0.2$; typical of the Milky Way disk) NS ULXs dominate the total population with an equal fraction of observed BH and NS ULXs. We note that, following the discussion above, the {\it observed} number of BH ULXs is always (in our simulated test cases) very similar to the total number of BH ULXs whilst the {\it observed} number of NS ULXs is always significantly below the total number of NS ULXs.



\subsection{Beamed vs. isotropic as a function of stellar population age}
\begin{figure*}[t]
    \centering
    \includegraphics[width=1.0\textwidth]{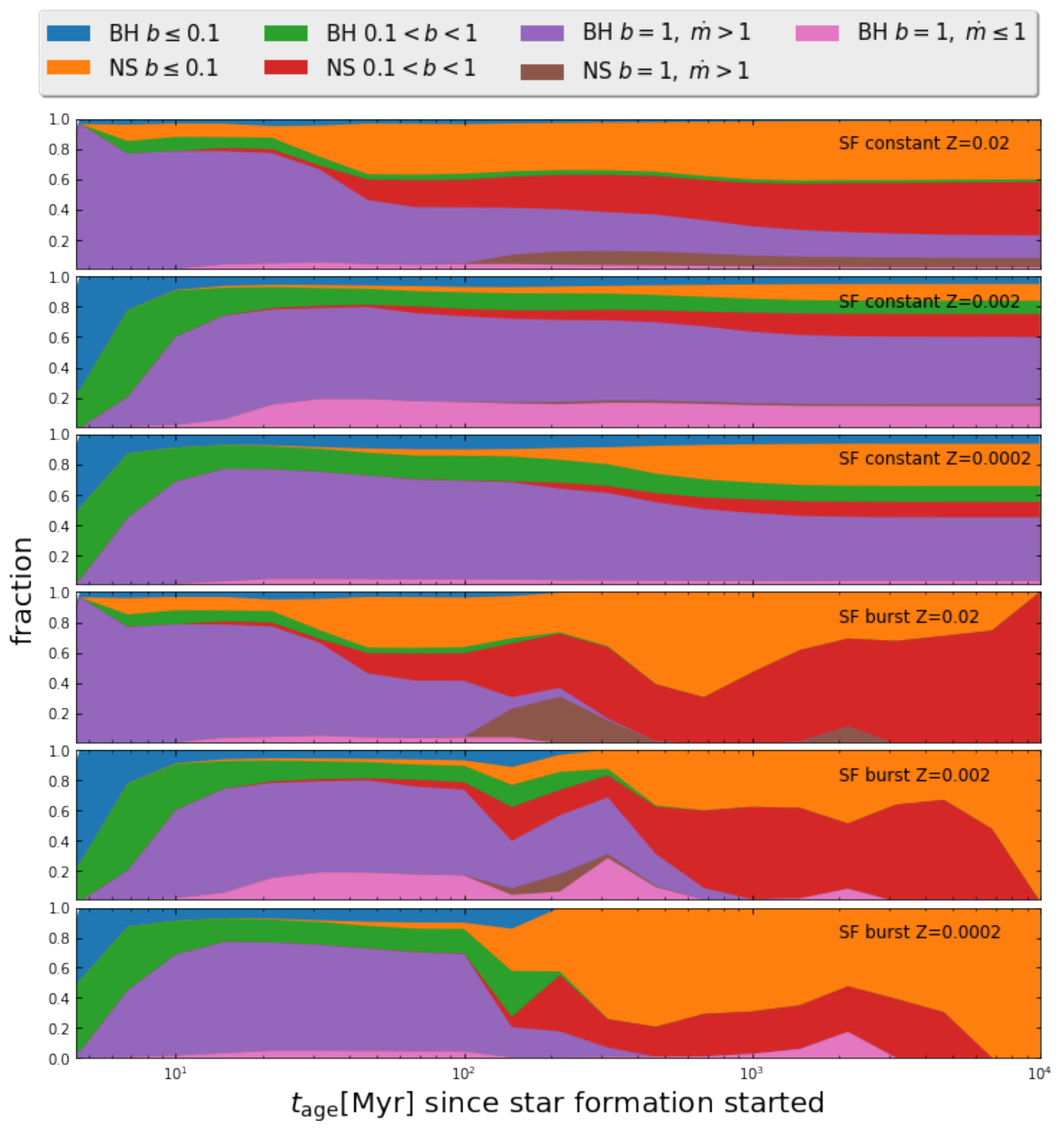}
    \caption{Fraction of BH/NS ULXs in the {\it total} population divided into four categories: sub-Eddington sources ($b=1$, $\dot{m}\leq1$),  isotropic sources ($b=1$, $\dot{m}<8.5$), mildly beamed sources ($1<b<0.1$), and highly beamed sources ($b\leq0.1$). Two star formation (SF) models are included: SF burst (duration $100\myr$; three upper plots) and constant SF (three lower plots). Three different metallicities are presented: $Z=\zsun=0.02$, $Z=10\%\zsun=0.002$, and $Z=1\%\zsun=0.0002$.  There are no NS ULXs in the sub-Eddington category. }
    \label{fig:bfrac}
\end{figure*}
\begin{figure*}[t]
    \centering
    \includegraphics[width=1.0\textwidth]{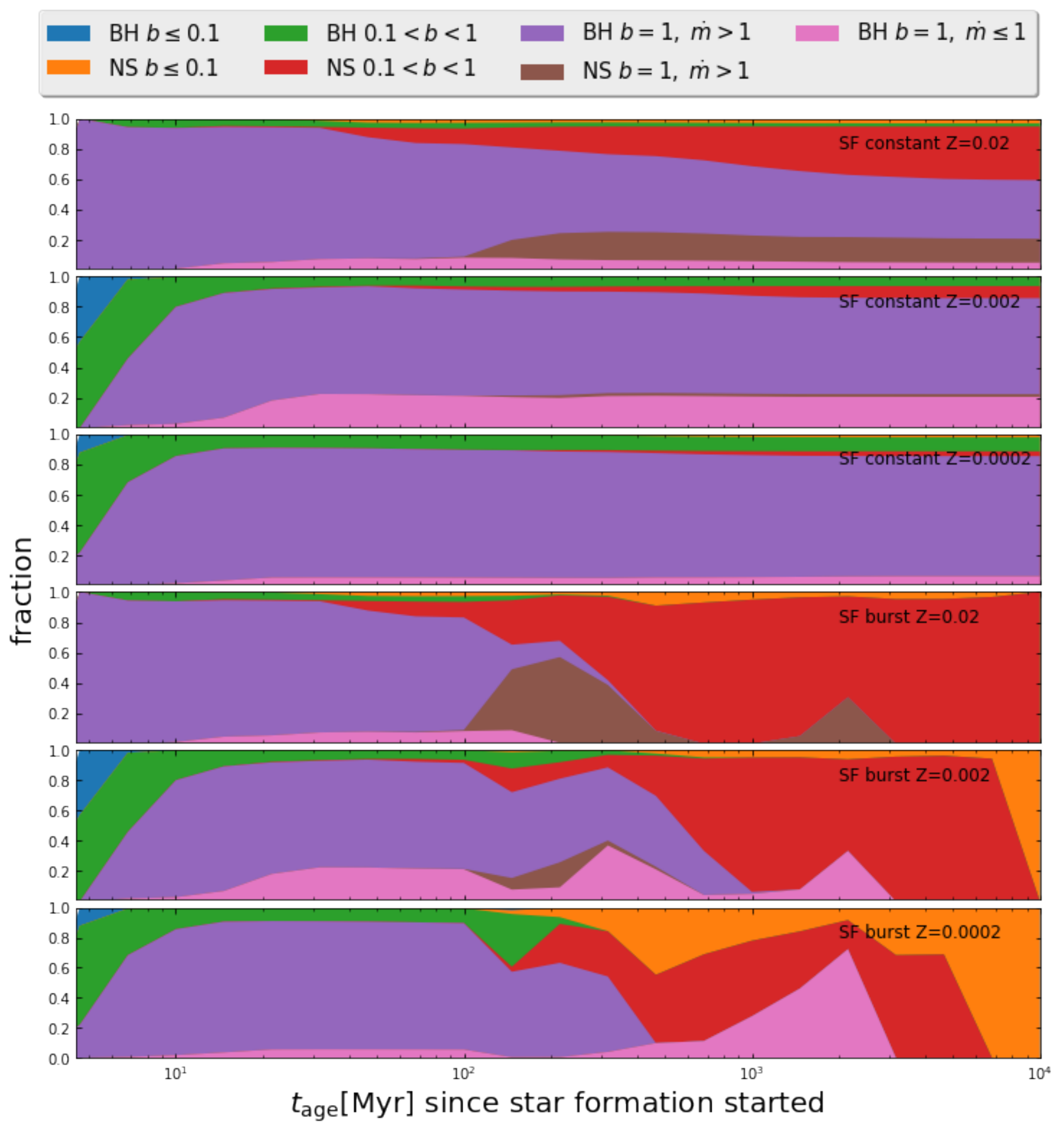}
    \caption{The same as Fig.~\ref{fig:bfrac}, but for the {\it observed} population of ULXs.}
    \label{fig:bfrac_obs}
\end{figure*}
%
%
%
%
%
%
%
%

Here we present a more detailed description of how the relative number of ULXs with different levels of beaming in the total population changes when the stellar population ages (depicted in Figure \ref{fig:bfrac}). 

Young ULXs ($t_{\rm age}\lesssim10\myr$) typically harbor BHs and massive ($\sim10\msun$) main-sequence (MS) donors which filled their Roche lobes (RL) due to nuclear evolution. However, the higher the metallicity, the stronger the mass loss in the stellar wind \citep[e.g.][]{Vink15}, therefore donors in a low metallicity environment are usually more massive and, as a consequence, usually provide a higher thermal time-scale MT rate (as this is proportional to $M_{don}$). Additionally,  low metallicity stars don't expand as significantly during the MS compared to solar metallicity stars \citep[e.g.][]{Pols9808}, so those which managed to fill their Roche lobes will tend to be more massive. As a result, for $Z<\zsun$, BHULXs are mostly significantly beamed ($b\leq0.1$), whereas for $Z=\zsun$ the emission is mainly isotropic. With time, the fraction of highly beamed BH ULXs quickly drops, and at an age of $\sim10\myr$ the total ULX population is dominated by mildly beamed ($1<b<0.1$) and isotropic BH ULXs.  A noticeable fraction of BH ULXs (up to $20\%$ \new{ of the total population} for sub solar metallicities, i.e. $Z=0.002$ and $Z=0.0002$ models) harbor a relatively massive BH with $M_{\rm BH}\gtrsim10\msun$, accreting at sub-Eddington rates. When the population is a few $10\myr$ old, highly and mildly beamed NS ULXs emerge and for  $Z=\zsun$ start to dominate the  total ULX population at $t_{\rm age}\approx100\myr$.
%
%
%
%
%
%
%
%
%
%
%
%
%

The continued evolution of the ULX population depends on the adopted SF history. In the case of constant SF (three lower plots on Fig.~\ref{fig:bfrac}). The fraction of  highly and mildly beamed sources (mostly NS ULXs) grows steadily to becomes nearly constant after $t_{\rm age}\approx1\gyr$. The fraction of beamed sources is then $\sim80\%$, $40\%$, $50\%$ for $Z=0.02$, $Z=0.002$, $Z=0.0002$, respectively. The fractions of highly and mildly beamed sources are comparable.  Highly beamed ULXs in an old stellar population ($t_{\rm age}>1\gyr$) are mostly NS ULXs.

A different situation occurs in post-burst populations when the SF is extinguished.  The number of BH ULXs drops quickly as no new BHs are produced and massive companions quickly evolve off the MS. NS ULXs, which are predominantly beamed, quickly become dominant and in a few $100\myr$ constitute nearly $100\%$ of all ULXs. Therefore, post-burst ULX populations are predicted to be mostly beamed, NS systems.

\begin{figure*}[t]
    \centering
    \includegraphics[width=1.0\textwidth]{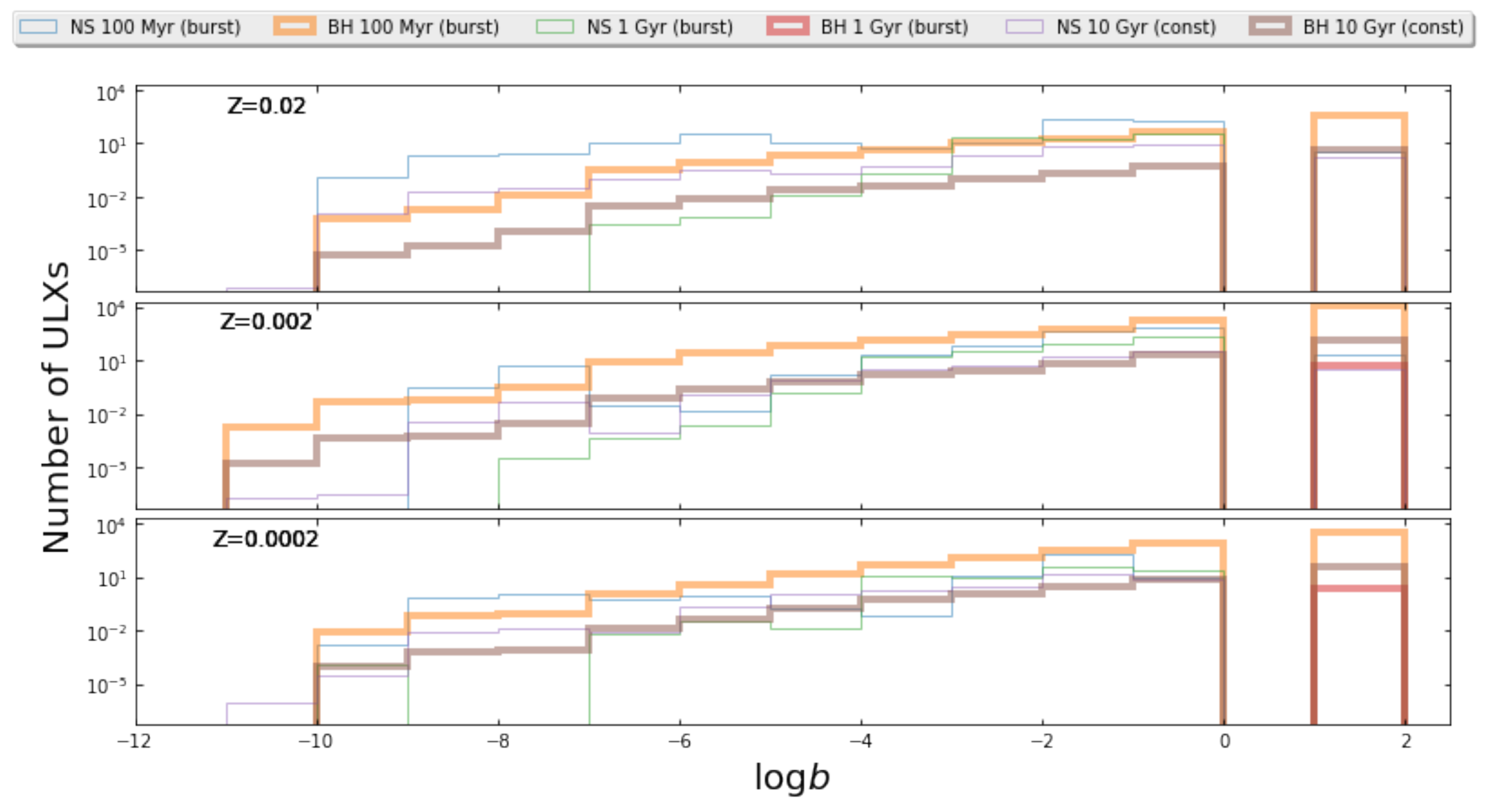}
    \caption{Histograms of beaming factor ($b$) in the total population of ULXs for three tested metallicities (marked in the upper left-hand corner) and three population ages: $100\myr$ (at the end of a SF burst), $1\gyr$ ($900 \myr$ after the end of a SF burst), and after $10\gyr$ of continuous SF. The rightmost peak (detached for the sake of clarity) represents isotropic emission ($b=1$; for both super- and sub-Eddington sources). Other bins represent beamed emission ($b<1$).
    }
    \label{fig:bdistr}
\end{figure*}
%
%
%
%
Figure \ref{fig:bdistr} presents the distributions of beaming factor in the total population of ULXs for different metallicities, accretor types, and three representative population ages. Two are for burst SF: $100\myr$  after the end of SF burst and $900 \myr$ after the end respectively. The third case represents the distribution for a population age of $10\gyr$ and continuous SF.  ULXs emitting isotropically were subtracted from the distribution and are presented in a detached bin. The left-most tail is formed by extremely beamed sources, which although very luminous, contribute little to volume-limited populations due to a very small observation probability (Eq.~\ref{eq:pobs}).




\subsubsection{The most highly beamed sources }
\label{sec:highb}
%
%
%
%


ULXs characterized by the highest beaming of their radiation are the most luminous systems in the population. Although BH ULXs are predominantly isotropic emitters (except very young, $t_{\rm age}\lesssim10\myr$, populations), some of them may exhibit strong beaming ($b<0.1$). Such systems have typically Hertzsprung gap (HG) donors \citep[in contrast to MS donors which are the typical  companions in BH ULXs in general;][]{Wiktorowicz1509,Wiktorowicz1709} with masses between $5$--$10\msun$. Typical BH masses of these ULXs are similar to those in isotropic ones ($\sim6$--$10\msun$) for all metallicities.

\section{Results II - Flux-limited surveys}
\label{sec:fluxLimitedSurvey}

Until now, we have considered the observational properties of ULXs that would be found in a volume limited survey. However, flux-limited surveys (e.g. the ROSAT all-sky survey and the forthcoming eROSITA all-sky survey) provide the broadest indication of the population of ULXs. Using our numerical results, we performed a comparison between simulated and analytical estimates \citep{Middleton1708} for the ratio of the number of NS ULXs to BH ULXs in idealized flux-limited observations, assuming that all ULXs directed towards the observer will be detectable and recognizable. Such observations will naturally be biased towards more strongly beamed, more luminous sources which can be detected out to larger distances. As a result, the ratio of beamed  to isotropic sources increases in comparison to volume-limited observations.
%
%
%
%
%
%
%
%

We assume that within large volumes ($D\gtrsim10$ Mpc$^{3}$) the distribution of stellar mass is homogeneous and that light travel time does not influence the results significantly. Additionally, we use a uniform ("toy") Universe models with the same metallicity and SFH at any place for an easier comparison of simulated and analytical estimates.
After such a simplification, the limiting distance for detecting a source can be expressed as $D_{\rm lim}=\sqrt{L_{\rm X, app} / 4\pi f_{\rm lim}}$, where $L_{\rm X, app}$ is the apparent luminosity of the source and $f_{\rm lim}$ is the limiting observable flux. Consequently, the volume within which the source will be observable is $V\propto L_{\rm X, app}^{3/2}$, where the scaling factor depends on $f_{\rm lim}$, which we assume to be the same for all sources (i.e. the conditions of a flux-limited survey). If we additionally, define the mean number density of stars as $n$, the probability of observing a particular source will be \citep[c.f.][]{Middleton1708}:
\begin{equation}\label{eq:PobsV}
		P\propto f_{\rm SFH}\cdot n\cdot b\cdot V\propto f_{\rm SFH}\cdot n\cdot b\cdot L_{\rm X, app}^{3/2}\cdot f_{\rm lim}^{-3/2},
\end{equation}
where $f_{\rm SFH}$ is the probability that the particular system will presently be in the ULX phase. More precisely:
\begin{equation}
f_{\rm SFH}=\frac{1}{\int_0^{10\gyr}{\rm SFH}(t^\prime)dt^\prime}\int_{t_{\rm age}-t}^{t_{\rm age}-t+dt}{\rm SFH}(t^\prime)dt^\prime,
\end{equation}
where $t_{\rm age}$ is the age of the population since star formation started, $t$ is the age of a given system (time since ZAMS) during the ULX phase, $dt$ is the length of  the ULX phase, and ${\rm SFH}(t^\prime)$ is the star formation history. For the simplified case, we defined ${\rm SFH}(t^\prime)$ as: 
%
%
%
%
\begin{equation}
	{\rm SFH}(t^\prime)=6\frac{\msun}{\yr},
\end{equation}
for constant SF and as:
\begin{equation}
	{\rm SFH}(t^\prime)=\left\{\begin{array}{cl}
600&t^\prime\leq100\myr\\
0&t^\prime>100\myr
\end{array}\right.\frac{\msun}{\yr},
\end{equation}
for burst SF. The estimated number of NS or BH ULXs ($N_{\rm NS/BH}$) may then be calculated from:
\begin{equation}\label{eq:E}
E(N_{\rm NS/BH})=\sum P_{\rm NS/BH}\propto\sum\limits_{\rm NS/BH\; ULXs} f_{\rm SFH}\cdot b\cdot L_{\rm app}^\frac{3}{2},
\end{equation}
where $P_{\rm NS/BH}$ is the probability of observation of a particular NS or BH ULX (Eq.~\ref{eq:PobsV}), and the summation is performed over entire ULX lifetime for all NS and BH ULXs.  The ratio of NS ULXs to BH ULXs is then:
\begin{equation}\label{eq:NStoBH}
\frac{E(N_{\rm NS})}{E(N_{\rm BH})}=\frac{\sum\limits_{\rm NS\; ULXs} f_{\rm SFH}\cdot b\cdot L_{\rm app}^\frac{3}{2}}{\sum\limits_{\rm BH\; ULXs} f_{\rm SFH}\cdot b\cdot L_{\rm app}^\frac{3}{2}}.
\end{equation}


%
%

\begin{figure}[t]
    \centering
    \includegraphics[width=1.0\columnwidth]{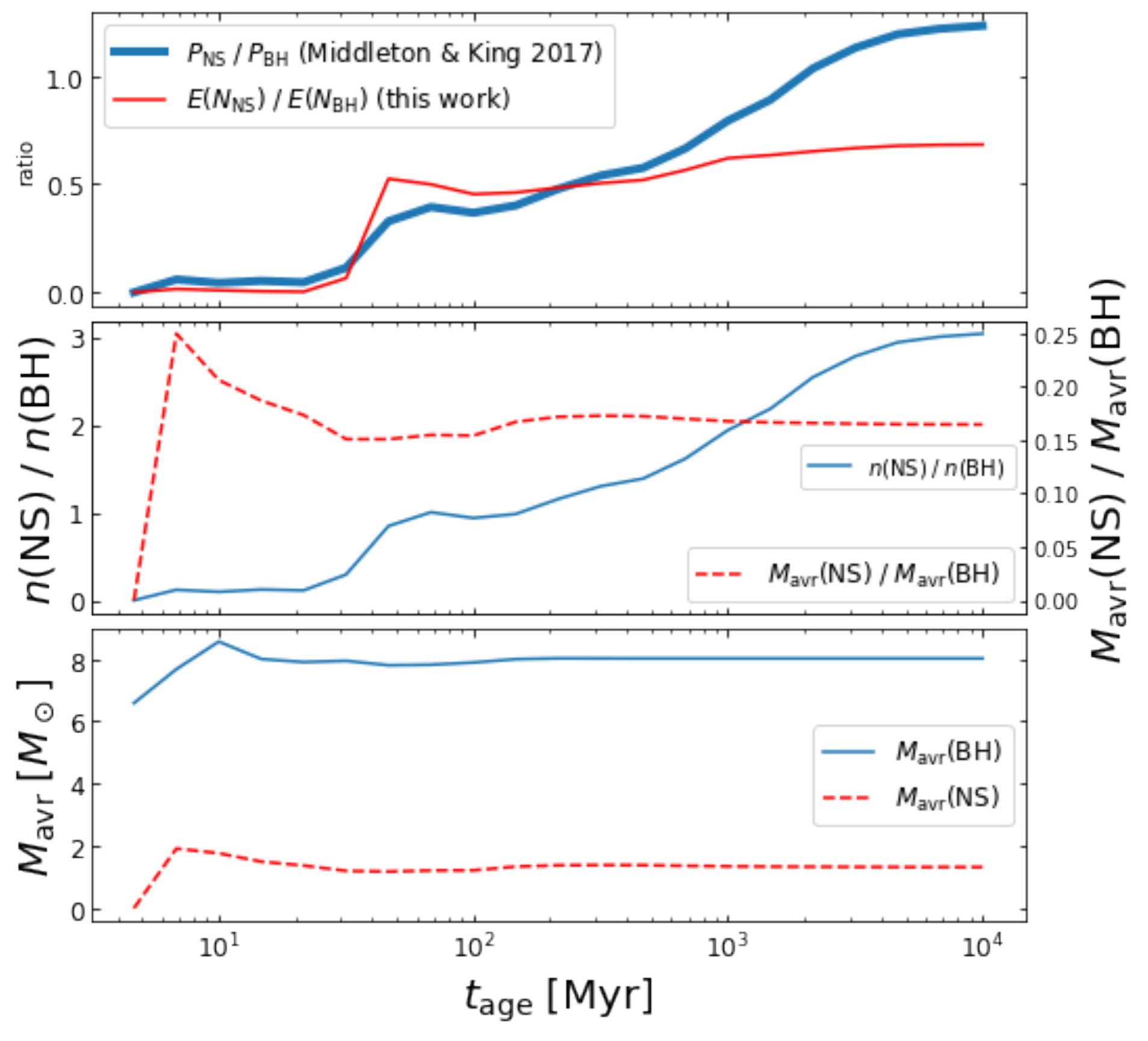}
    \caption{The upper panel shows the estimated ratio of NS ULXs to BH ULXs for our simulations ($E(N_{\rm NS})/E(N_{\rm BH})$; Eq.~\ref{eq:NStoBH}) and the analytic formula of \citet[$P_{\rm NS}/P_{\rm BH}$; Eq.~6]{Middleton1708}. Solar metallicity ($Z=0.02$) and constant SF through the last $10$ Gyr  were assumed. The middle and bottom panels present the number densities of NS/BH ULXs ($n({\rm NS}/{\rm BH})$; middle panel), ratio of average masses of NS/BH accretors ($M_{\rm avr}({\rm NS}/{\rm BH})$; middle panel), and average masses of NS/BH separately (bottom panel) obtained from our simulations and used for the calculation of $P_{\rm NS}/P_{\rm BH}$.  }
    \label{fig:matt}
\end{figure}
\begin{figure}[t]
    \centering
    \includegraphics[width=1.0\columnwidth]{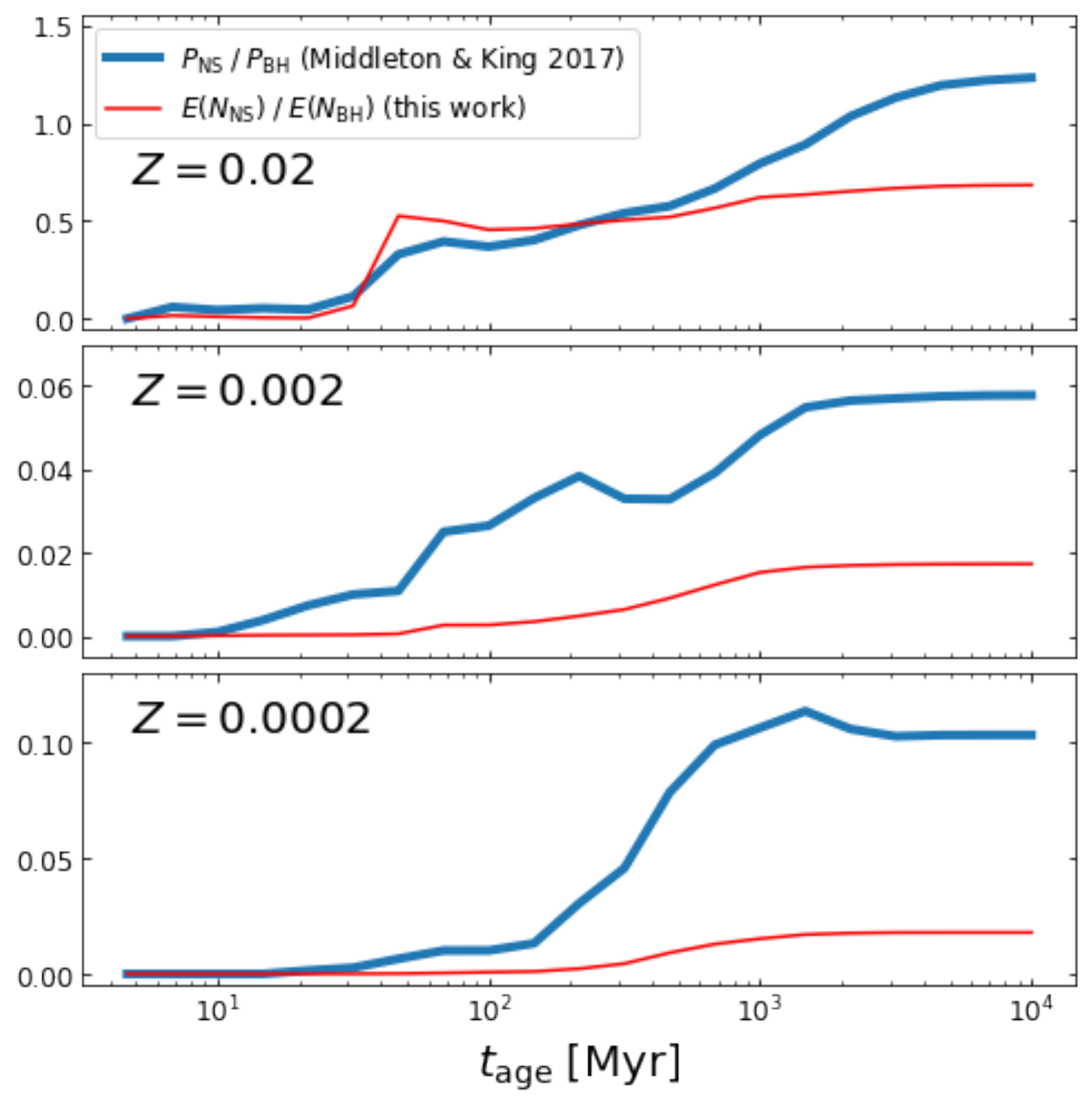}
    \caption{Same as Fig.~\ref{fig:matt}, upper panel, but for different metallicities. For lower metallicities there are relatively far fewer NS ULXs (both for our results and using the \citet{Middleton1708} prescription) than for solar metallicity ($Z=0.02$).}
    \label{fig:mattZ}
\end{figure}


In Fig.~\ref{fig:matt} we show the comparison of our simulated estimate of the relative number of NS ULXs to BH ULXs (Eq.~\ref{eq:NStoBH}), to the simplified analytical formula of \citet[Eq.~6]{Middleton1708}. The number densities and mean masses of accretors  enter into their formula and we calculate these directly from our simulations (and are also provided in Fig.~\ref{fig:matt}). Clearly the results are consistent at $t_{age} < 1~$Gyr and only differ across the entire simulation by a factor $\lesssim$ 2; this is a clear validation of the simplified approach by \citet{Middleton1708}. We note that, when the various contributions to the overall population are considered, our results diverge to a greater extent at low metallicities, and at late times can differ by up to an order of magnitude for $Z = 0.0002$ (see Fig.~\ref{fig:mattZ}). There are two reasons for this discrepancy, firstly, \citet{Middleton1708} assumed that $\dot{m}_{\rm tr}$ ($\dot{m}_0$ in their work) is always $\geq1$ (see their Eq.~1), whereas we consider BH ULXs which can be classified as ULXs with $\dot{m}_{\rm tr} < 1$ and in the population are mostly unbeamed. Secondly, in obtaining their Eq.~4 it was assumed that MT rates for NS ULXs and BH ULXs are similar whereas this is not always true because BH ULXs can achieve much higher stable MT rates (e.g. from massive stars, $M_{\rm donor}>10\msun$), which would otherwise lead to a dynamical instability for NS accretors. These assumptions contribute to the discrepancy the most where the metallicity of the environment is lowest, as masses of BHs are then higher on average, so there are more ULXs emitting isotropically.
\section{Discussion}

Our simulations should allow us to answer the fundamental question, is beaming necessary to explain the observed population of ULXs? Unfortunately, without detailed information on the environment in which observed ULXs reside, such as SFH and metallicity distribution, we are unable to make reliable predictions. However, the overall results from our simulations regarding the populations of NS and BH ULXs and the role of beaming should be robust to different evolutionary models \citep{Wiktorowicz1709} and these are important for understanding the observed population in both volume and flux-limited surveys. 

There are two key results from our simulations: the fraction of observed to total sample of ULXs and the ratio of NS vs BH ULXs. In general, for star forming regions, the fraction of observed to the total sample of ULXs is $\sim0.8$ (independently of metallicty) due to a high abundance of BH ULXs which typically emit isotropically, except for the very early ages ($t_{\rm age}<10\myr$) when most of the BH ULX are singificantly beamed and the fraction of observed to total sample is smaller. Conversely, for a solar metallicity environment after a long SF episode ($\gtrsim100\myr$ of continuous star formation), the observed population is dominated by NS ULXs. For old stellar populations where the SF ceased $\sim1\gyr$ ago, the ratio of observed to total population of ULXs is typically $\sim0.2$  because it consists nearly exclusively of NS ULXs, which are nearly always beamed. 
%
%
%
%
%
%

As we show in Fig.~\ref{fig:num}, the relative fraction of NS to BH ULXs changes as a function of both metallicty, star formation model (continuous or burst) and age since star formation commenced. In the case of continuous star formation, BH ULXs dominate the observed population of ULXs (the lowest ratio to NS ULXs is 1:1 for solar metalicities at late times). For burst star-formation, young ULX populations are dominated by BH ULXs but this changes as the population ages and, post star-formation, NS ULXs dominate both the observed and total population of ULXs.

Our adopted geometrical beaming model predicts extremely strong beaming for some ULXs (Fig.~\ref{fig:bdistr}). The corresponding apparent luminosities are well above those observed for extreme ULXs ($L_{\rm X,max}\approx1\times10^{42}\ergs$). To avoid such a situation, in the previous work \citep{Wiktorowicz1709}, we applied a saturation threshold for beaming at $b_{\rm lim}=3.2\times10^{-3}$ ($\dot{m}_{\rm lim}\approx150$), which capped the luminosities to $L_{\rm X}\lesssim10^{42}\ergs$. In this paper we found that adopting the saturation does not change the results and conclusions significantly.

\begin{figure}[t]
    \centering
    \includegraphics[width=1.0\columnwidth]{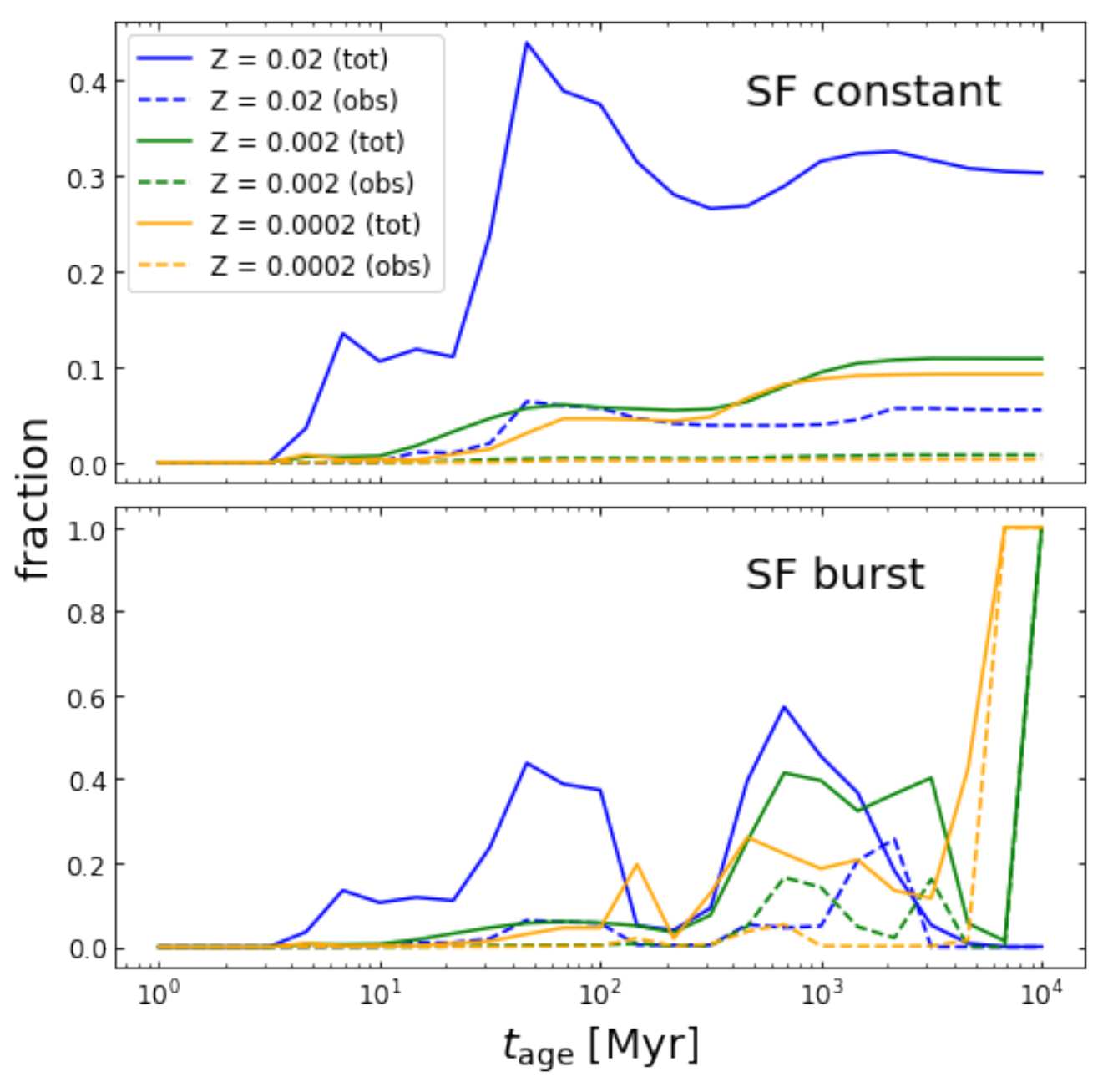}
    \caption{Fraction of ULXs (with both BH and NS accretors) undergoing thermal timescale mass transfer in the total (tot; continuous lines) and observed (obs; dashed lines) population. Two star formation (SF) models are included: constant SF (upper panel) and SF burst (duration $100\myr$; lower panel). Three different metallicities are presented: $Z=\zsun=0.02$, $Z=10\%\zsun=0.002$, and $Z=1\%\zsun=0.0002$.}
    \label{fig:mt4frac}
\end{figure}

\new{\citet{King0105} suggested that the thermal-timescale mass transfer is the best process to fuel a ULX. Indeed, part of the ULXs in our results, including the most luminous ones, experience a thermal timescale mass transfer. In these sources, the donor is not in thermal equilibrium and its characteristics (especially the mass-radius relation) may differ strongly from those calculated for thermally stable stars, which are used in population synthesis codes. Consequently, the predicted observational properties of ULXs (e.g. the duration of the ULX phase, or peak luminosity) may differ in contrast to the results of detailed codes \citep[e.g.][find a difference by a factor of $\sim2$ in the duration of the RLOF phase of the $1.3\msun$ NS and $1.6\msun$ MS donor]{Belczynski0801}. However, in our results, the observed and total populations are dominated by ULXs undergoing nuclear timescale mass transfer (exception are very old populations; see Fig.~\ref{fig:mt4frac}). It is the direct consequence of a much longer duration of the mass transfer phases. In spite of the problems with calculating the realistic mass transfer rates and durations of thermal timescale mass transfer, we show that it can affect our results only for old stellar populations, whereas ULXs are mostly found in star- forming environments \citep[e.g.][]{Gao0310,Fabbiano0307}.

Specificly, we found that typically a BH ULX (except the most luminous ones with $L_\mathrm{X}\gtrsim10^{41}\ergs$), undergo a nuclear timescale mass transfer. Therefore, ULX populations dominated by BH accretors are, consequently, dominated by nuclear timescale mass transfer (typically the fraction is $\gtrsim90\%$ for both total and observed populations). As we predict that BH-dominated ULX populations are typical for star formation bursts and constant star formation environments (except the total population in prolonged constant star formation in solar metallicity environment; see Fig.~\ref{fig:num}), where most of the ULXs are found \citep[e.g.,][]{Swartz0410,Swartz0909}, also the majority of ULXs are expected to undergo a mass transfer on a nuclear timescale.

NS ULXs may have much higher fraction of systems undergoing a thermal timescale mass transfer which depends on SFH (see Fig.~\ref{fig:mt4frac}). For solar metallicity environment, the fraction of thermal timescale mass transfer among ULXs reaches $\sim40\%$ ($\sim5\%$ for observed population) mostly due to a high fraction of NS accretors. Also in older environ entsm(a few $100\myr$ after the ceasation of star formation) the population is dominated by NS ULX, thus the fraction of thermal timescale mass transfer is higher (up to $\sim50\%$ for total poulation of up to $\sim20\%$ for observed population). We note that the fraction reaches $100\%$ for very old stellar populations ($t_\mathrm{age}\gtrsim9\gyr$) in low-metallicity environments ($Z=0.002$ and $Z=0.0002$), but it may be a result of very limited statistics.}


\section{Summary}

In this paper, we have analyzed how geometrical beaming, i.e. anisotropic emission of radiation, affects the observed sample of ULXs when compared to the total sample of these objects, some part of which is hidden from our view. Our simulated results are based on the previous analysis of ULX populations in different environments published in \citet{Wiktorowicz1709} and may be seen as an extension of the previous work. The utilized beaming model (beaming factor $b\propto \dot{m}^{-2}$) is based on theoretical and observational grounds \citep{King0902}.

We show that ULXs harboring BH accretors are typically emitting isotropically ($b=1$)\new{ and undergo a nuclear timescale mass transfer}, whereas those with NS accretors are predominantly beamed (typically $b=0.07-0.2$)\new{ and in most cases mass transfer occurs on a thermal timescale}. Our analysis shows that the beaming is dependent on different stellar environments; very young (burst) populations ($t_{\rm age}<10\myr$), dominated by BH ULXs, are significantly beamed whilst BH ULXs in older stellar populations are usually isotropic emitters. However, the majority of NS ULXs are {\it always} beamed, irrespective of stellar environment. In terms of the relative ratio of species, we find that the ratio of NS ULXs to BH ULXs is higher in the total sample than in the observed sample. In the case of continuous star-formation, BH ULXs typically outnumber the NS ULXs in the observed sample. Whilst BH ULXs also outnumber the NS ULXs in the observed sample for burst star-formation at early times, post star formation, NS ULXs tend to dominate the observed population instead. In the case of the latter, the observed NS ULXs represent only 20\% of the total NS ULX population and many are expected to be obscured from view \citep[in the absence of precession which may act to bring some into view - see][]{Dauser1704,Middleton1810}.



Finally, we found that the ratio of the number of NS ULXs to BH ULXs in idealized flux-limited observations is consistent within a factor $\lesssim 2$ of that found by \citet{Middleton1708}, with divergence at late times and lower metallicities where, in the case of the latter, large MT rates can lead to instabilities for accreting NS systems and very massive BHs are more common.


\acknowledgements
\new{We are grateful to the anonymous referee who help us to improve the paper.} JPL and MM thank Andrew King for many inspiring discussions.
We are thankful to thousends of volunteers, who took part in the {\it Universe@Home} project\footnote{\url{https://universeathome.pl/}} and provided their computers for this research.
GW is partly supported by the President’s International Fellowship Initiative (PIFI) of the Chinese Academy of Sciences under grant no.2018PM0017 and by the Strategic Priority Research Program of the Chinese Academy of Science Multi-waveband Gravitational Wave Universe (Grant No. XDB23040000). 
JPL \& KB acknowledge support by the National Science Centre, Poland grant 2015/19/B/ST9/01099 and JPL by a grant from the French Space Agency CNES.
MM thanks STFC for support via an Ernest Rutherford Fellowship.

\bibliographystyle{apj}
\bibliography{ms}

\begin{thebibliography}{}
\expandafter\ifx\csname natexlab\endcsname\relax\def\natexlab#1{#1}\fi

\bibitem[{{Abarca} {et~al.}(2018){Abarca}, {Klu{\'z}niak}, \& {S{\c
  a}dowski}}]{Abarca1809}
{Abarca}, D., {Klu{\'z}niak}, W., \& {S{\c a}dowski}, A. 2018, \mnras, 479,
  3936

\bibitem[{{Abbott} {et~al.}(2016){Abbott}, {Abbott}, {Abbott}, {Abernathy},
  {Acernese}, {Ackley}, {Adams}, {Adams}, {Addesso}, {Adhikari}, \&
  et~al.}]{Abbott1602}
{Abbott}, B.~P., {Abbott}, R., {Abbott}, T.~D., {et~al.} 2016, \apjl, 818, L22

\bibitem[{{Bachetti} {et~al.}(2014){Bachetti}, {Harrison}, {Walton},
  {Grefenstette}, {Chakrabarty}, {F{\"u}rst}, {Barret}, {Beloborodov}, {Boggs},
  {Christensen}, {Craig}, {Fabian}, {Hailey}, {Hornschemeier}, {Kaspi},
  {Kulkarni}, {Maccarone}, {Miller}, {Rana}, {Stern}, {Tendulkar}, {Tomsick},
  {Webb}, \& {Zhang}}]{Bachetti1410}
{Bachetti}, M., {Harrison}, F.~A., {Walton}, D.~J., {et~al.} 2014, \nat, 514,
  202

\bibitem[{{Basko} \& {Sunyaev}(1976)}]{Basko7605}
{Basko}, M.~M., \& {Sunyaev}, R.~A. 1976, \mnras, 175, 395

\bibitem[{{Begelman} {et~al.}(2006){Begelman}, {King}, \&
  {Pringle}}]{Begelman0607}
{Begelman}, M.~C., {King}, A.~R., \& {Pringle}, J.~E. 2006, \mnras, 370, 399

\bibitem[{{Belczynski} {et~al.}(2002){Belczynski}, {Kalogera}, \&
  {Bulik}}]{Belczynski0206}
{Belczynski}, K., {Kalogera}, V., \& {Bulik}, T. 2002, \apj, 572, 407

\bibitem[{{Belczynski} {et~al.}(2008){Belczynski}, {Kalogera}, {Rasio}, {Taam},
  {Zezas}, {Bulik}, {Maccarone}, \& {Ivanova}}]{Belczynski0801}
{Belczynski}, K., {Kalogera}, V., {Rasio}, F.~A., {et~al.} 2008, \apjs, 174,
  223

\bibitem[{{Belczynski} \& {Taam}(2004)}]{Belczynski0403}
{Belczynski}, K., \& {Taam}, R.~E. 2004, \apj, 603, 690

\bibitem[{{Belczynski} {et~al.}(2012){Belczynski}, {Wiktorowicz}, {Fryer},
  {Holz}, \& {Kalogera}}]{Belczynski1209}
{Belczynski}, K., {Wiktorowicz}, G., {Fryer}, C.~L., {Holz}, D.~E., \&
  {Kalogera}, V. 2012, \apj, 757, 91

\bibitem[{{Bovy} \& {Rix}(2013)}]{Bovy1312}
{Bovy}, J., \& {Rix}, H.-W. 2013, \apj, 779, 115

\bibitem[{{Brightman} {et~al.}(2018){Brightman}, {Harrison}, {F{\"u}rst},
  {Middleton}, {Walton}, {Stern}, {Fabian}, {Heida}, {Barret}, \&
  {Bachetti}}]{Brightman1804}
{Brightman}, M., {Harrison}, F.~A., {F{\"u}rst}, F., {et~al.} 2018, Nature
  Astronomy, 2, 312

\bibitem[{{Carpano} {et~al.}(2018){Carpano}, {Haberl}, {Maitra}, \&
  {Vasilopoulos}}]{Carpano1805}
{Carpano}, S., {Haberl}, F., {Maitra}, C., \& {Vasilopoulos}, G. 2018, \mnras,
  476, L45

\bibitem[{{Clark} {et~al.}(2014){Clark}, {Ritchie}, {Najarro}, {Langer}, \&
  {Negueruela}}]{Clark0514}
{Clark}, J.~S., {Ritchie}, B.~W., {Najarro}, F., {Langer}, N., \& {Negueruela},
  I. 2014, \aap, 565, A90

\bibitem[{{Colbert} \& {Mushotzky}(1999)}]{Colbert9907}
{Colbert}, E.~J.~M., \& {Mushotzky}, R.~F. 1999, \apj, 519, 89

\bibitem[{{Dai} {et~al.}(2018){Dai}, {McKinney}, {Roth}, {Ramirez-Ruiz}, \&
  {Miller}}]{Dai1806}
{Dai}, L., {McKinney}, J.~C., {Roth}, N., {Ramirez-Ruiz}, E., \& {Miller},
  M.~C. 2018, \apjl, 859, L20

\bibitem[{{Dauser} {et~al.}(2017){Dauser}, {Middleton}, \&
  {Wilms}}]{Dauser1704}
{Dauser}, T., {Middleton}, M., \& {Wilms}, J. 2017, \mnras, 466, 2236

\bibitem[{{de Mink} \& {Belczynski}(2015)}]{deMink1511}
{de Mink}, S.~E., \& {Belczynski}, K. 2015, \apj, 814, 58

\bibitem[{{Fabbiano} {et~al.}(2003){Fabbiano}, {King}, {Zezas}, {Ponman},
  {Rots}, \& {Schweizer}}]{Fabbiano0307}
{Fabbiano}, G., {King}, A.~R., {Zezas}, A., {et~al.} 2003, \apj, 591, 843

\bibitem[{{Fabrika}(2004)}]{Fabrika04}
{Fabrika}, S. 2004, Astrophysics and Space Physics Reviews, 12, 1

\bibitem[{{Finke} \& {Razzaque}(2017)}]{Finke1712}
{Finke}, J.~D., \& {Razzaque}, S. 2017, \mnras, 472, 3683

\bibitem[{{Fragos} {et~al.}(2015){Fragos}, {Linden}, {Kalogera}, \&
  {Sklias}}]{Fragos1503}
{Fragos}, T., {Linden}, T., {Kalogera}, V., \& {Sklias}, P. 2015, \apjl, 802,
  L5

\bibitem[{{Fryer} {et~al.}(2012){Fryer}, {Belczynski}, {Wiktorowicz},
  {Dominik}, {Kalogera}, \& {Holz}}]{Fryer1204}
{Fryer}, C.~L., {Belczynski}, K., {Wiktorowicz}, G., {et~al.} 2012, \apj, 749,
  91

\bibitem[{{F{\"u}rst} {et~al.}(2016){F{\"u}rst}, {Walton}, {Harrison}, {Stern},
  {Barret}, {Brightman}, {Fabian}, {Grefenstette}, {Madsen}, {Middleton},
  {Miller}, {Pottschmidt}, {Ptak}, {Rana}, \& {Webb}}]{Furst1611}
{F{\"u}rst}, F., {Walton}, D.~J., {Harrison}, F.~A., {et~al.} 2016, \apjl, 831,
  L14

\bibitem[{{Gao} {et~al.}(2003){Gao}, {Wang}, {Appleton}, \& {Lucas}}]{Gao0310}
{Gao}, Y., {Wang}, Q.~D., {Appleton}, P.~N., \& {Lucas}, R.~A. 2003, \apjl,
  596, L171

\bibitem[{{Heida} {et~al.}(2016){Heida}, {Jonker}, {Torres}, {Roberts},
  {Walton}, {Moon}, {Stern}, \& {Harrison}}]{Heida1606}
{Heida}, M., {Jonker}, P.~G., {Torres}, M.~A.~P., {et~al.} 2016, \mnras, 459,
  771

\bibitem[{{Heida} {et~al.}(2014){Heida}, {Jonker}, {Torres}, {Kool},
  {Servillat}, {Roberts}, {Groot}, {Walton}, {Moon}, \& {Harrison}}]{Heida1408}
---. 2014, \mnras, 442, 1054

\bibitem[{{Hobbs} {et~al.}(2005){Hobbs}, {Lorimer}, {Lyne}, \&
  {Kramer}}]{Hobbs0507}
{Hobbs}, G., {Lorimer}, D.~R., {Lyne}, A.~G., \& {Kramer}, M. 2005, \mnras,
  360, 974

\bibitem[{{Israel} {et~al.}(2017){Israel}, {Belfiore}, {Stella}, {Esposito},
  {Casella}, {De Luca}, {Marelli}, {Papitto}, {Perri}, {Puccetti}, {Castillo},
  {Salvetti}, {Tiengo}, {Zampieri}, {D'Agostino}, {Greiner}, {Haberl},
  {Novara}, {Salvaterra}, {Turolla}, {Watson}, {Wilms}, \&
  {Wolter}}]{Israel1609}
{Israel}, G.~L., {Belfiore}, A., {Stella}, L., {et~al.} 2017, Science, 355, 817

\bibitem[{{Ivanova} {et~al.}(2013){Ivanova}, {Justham}, {Chen}, {De Marco},
  {Fryer}, {Gaburov}, {Ge}, {Glebbeek}, {Han}, {Li}, {Lu}, {Marsh},
  {Podsiadlowski}, {Potter}, {Soker}, {Taam}, {Tauris}, {van den Heuvel}, \&
  {Webbink}}]{Ivanova1302}
{Ivanova}, N., {Justham}, S., {Chen}, X., {et~al.} 2013, \aapr, 21, 59

\bibitem[{{Kaaret} {et~al.}(2017){Kaaret}, {Feng}, \& {Roberts}}]{Kaaret1708}
{Kaaret}, P., {Feng}, H., \& {Roberts}, T.~P. 2017, \araa, 55, 303

\bibitem[{{Kaaret} {et~al.}(2004){Kaaret}, {Ward}, \& {Zezas}}]{Kaaret0407}
{Kaaret}, P., {Ward}, M.~J., \& {Zezas}, A. 2004, \mnras, 351, L83

\bibitem[{{King}(2011)}]{King1105}
{King}, A. 2011, \apjl, 732, L28

\bibitem[{{King} \& {Lasota}(2014)}]{King1410}
{King}, A., \& {Lasota}, J.-P. 2014, \mnras, 444, L30

\bibitem[{{King} \& {Lasota}(2016)}]{King1605}
---. 2016, \mnras, 458, L10

\bibitem[{{King} \& {Lasota}(2019)}]{King1905}
---. 2019, \mnras, 485, 3588

\bibitem[{{King} {et~al.}(2017){King}, {Lasota}, \& {Klu{\'z}niak}}]{King1702}
{King}, A., {Lasota}, J.-P., \& {Klu{\'z}niak}, W. 2017, \mnras, 468, L59

\bibitem[{{King}(2009)}]{King0902}
{King}, A.~R. 2009, \mnras, 393, L41

\bibitem[{{King} {et~al.}(2001){King}, {Davies}, {Ward}, {Fabbiano}, \&
  {Elvis}}]{King0105}
{King}, A.~R., {Davies}, M.~B., {Ward}, M.~J., {Fabbiano}, G., \& {Elvis}, M.
  2001, \apjl, 552, L109

\bibitem[{{King} \& {Ritter}(1999)}]{King9910}
{King}, A.~R., \& {Ritter}, H. 1999, \mnras, 309, 253

\bibitem[{{Klencki} {et~al.}(2018){Klencki}, {Moe}, {Gladysz}, {Chruslinska},
  {Holz}, \& {Belczynski}}]{Klencki1811}
{Klencki}, J., {Moe}, M., {Gladysz}, W., {et~al.} 2018, \aap, 619, A77

\bibitem[{{Koliopanos} {et~al.}(2019){Koliopanos}, {Vasilopoulos}, {Buchner},
  {Maitra}, \& {Haberl}}]{Koliopanos1901}
{Koliopanos}, F., {Vasilopoulos}, G., {Buchner}, J., {Maitra}, C., \& {Haberl},
  F. 2019, \aap, 621, A118

\bibitem[{{Koliopanos} {et~al.}(2017){Koliopanos}, {Vasilopoulos}, {Godet},
  {Bachetti}, {Webb}, \& {Barret}}]{Koliopanos1712}
{Koliopanos}, F., {Vasilopoulos}, G., {Godet}, O., {et~al.} 2017, \aap, 608,
  A47

\bibitem[{{Kroupa} \& {Weidner}(2003)}]{Kroupa0312}
{Kroupa}, P., \& {Weidner}, C. 2003, \apj, 598, 1076

\bibitem[{{Lasota}(2015)}]{Lasota1505}
{Lasota}, J.-P. 2015, ArXiv e-prints, arXiv:1505.02172

\bibitem[{{Lasota} {et~al.}(2015){Lasota}, {King}, \& {Dubus}}]{Lasota1503}
{Lasota}, J.-P., {King}, A.~R., \& {Dubus}, G. 2015, \apjl, 801, L4

\bibitem[{{Lasota} {et~al.}(2016){Lasota}, {Vieira}, {Sadowski}, {Narayan}, \&
  {Abramowicz}}]{Lasota1603}
{Lasota}, J.-P., {Vieira}, R.~S.~S., {Sadowski}, A., {Narayan}, R., \&
  {Abramowicz}, M.~A. 2016, \aap, 587, A13

\bibitem[{{Licquia} \& {Newman}(2015)}]{Licquia1506}
{Licquia}, T.~C., \& {Newman}, J.~A. 2015, \apj, 806, 96

\bibitem[{{Linden} {et~al.}(2010){Linden}, {Kalogera}, {Sepinsky}, {Prestwich},
  {Zezas}, \& {Gallagher}}]{Linden1012}
{Linden}, T., {Kalogera}, V., {Sepinsky}, J.~F., {et~al.} 2010, \apj, 725, 1984

\bibitem[{{Liu} {et~al.}(2004){Liu}, {Bregman}, \& {Seitzer}}]{Liu0402}
{Liu}, J.-F., {Bregman}, J.~N., \& {Seitzer}, P. 2004, \apj, 602, 249

\bibitem[{{Maciel} {et~al.}(2012){Maciel}, {Rocha-Pinto}, \&
  {Costa}}]{Maciel1208}
{Maciel}, W.~J., {Rocha-Pinto}, H.~J., \& {Costa}, R.~D.~D. 2012, in IAU
  Symposium, Vol. 284, The Spectral Energy Distribution of Galaxies - SED 2011,
  ed. R.~J. {Tuffs} \& C.~C. {Popescu}, 379--381

\bibitem[{{Madhusudhan} {et~al.}(2006){Madhusudhan}, {Justham}, {Nelson},
  {Paxton}, {Pfahl}, {Podsiadlowski}, \& {Rappaport}}]{Madhusudhan0604}
{Madhusudhan}, N., {Justham}, S., {Nelson}, L., {et~al.} 2006, \apj, 640, 918

\bibitem[{{Madhusudhan} {et~al.}(2008){Madhusudhan}, {Rappaport},
  {Podsiadlowski}, \& {Nelson}}]{Madhusudhan0812}
{Madhusudhan}, N., {Rappaport}, S., {Podsiadlowski}, P., \& {Nelson}, L. 2008,
  \apj, 688, 1235

\bibitem[{{Marchant} {et~al.}(2017){Marchant}, {Langer}, {Podsiadlowski},
  {Tauris}, {de Mink}, {Mandel}, \& {Moriya}}]{Marchant1708}
{Marchant}, P., {Langer}, N., {Podsiadlowski}, P., {et~al.} 2017, \aap, 604,
  A55

\bibitem[{{Medvedev} \& {Fabrika}(2010)}]{Medvedev1002}
{Medvedev}, A., \& {Fabrika}, S. 2010, \mnras, 402, 479

\bibitem[{{Middleton} {et~al.}(2015{\natexlab{a}}){Middleton}, {Heil},
  {Pintore}, {Walton}, \& {Roberts}}]{Middleton1503}
{Middleton}, M.~J., {Heil}, L., {Pintore}, F., {Walton}, D.~J., \& {Roberts},
  T.~P. 2015{\natexlab{a}}, \mnras, 447, 3243

\bibitem[{{Middleton} \& {King}(2016)}]{Middleton1610}
{Middleton}, M.~J., \& {King}, A. 2016, \mnras, 462, L71

\bibitem[{{Middleton} \& {King}(2017)}]{Middleton1708}
---. 2017, \mnras, 470, L69

\bibitem[{{Middleton} {et~al.}(2012){Middleton}, {Sutton}, {Roberts},
  {Jackson}, \& {Done}}]{Middleton1203}
{Middleton}, M.~J., {Sutton}, A.~D., {Roberts}, T.~P., {Jackson}, F.~E., \&
  {Done}, C. 2012, \mnras, 420, 2969

\bibitem[{{Middleton} {et~al.}(2015{\natexlab{b}}){Middleton}, {Walton},
  {Fabian}, {Roberts}, {Heil}, {Pinto}, {Anderson}, \&
  {Sutton}}]{Middleton1512}
{Middleton}, M.~J., {Walton}, D.~J., {Fabian}, A., {et~al.} 2015{\natexlab{b}},
  \mnras, 454, 3134

\bibitem[{{Middleton} {et~al.}(2013){Middleton}, {Miller-Jones}, {Markoff},
  {Fender}, {Henze}, {Hurley-Walker}, {Scaife}, {Roberts}, {Walton},
  {Carpenter}, {Macquart}, {Bower}, {Gurwell}, {Pietsch}, {Haberl}, {Harris},
  {Daniel}, {Miah}, {Done}, {Morgan}, {Dickinson}, {Charles}, {Burwitz}, {Della
  Valle}, {Freyberg}, {Greiner}, {Hernanz}, {Hartmann}, {Hatzidimitriou},
  {Riffeser}, {Sala}, {Seitz}, {Reig}, {Rau}, {Orio}, {Titterington}, \&
  {Grainge}}]{Middleton1301}
{Middleton}, M.~J., {Miller-Jones}, J.~C.~A., {Markoff}, S., {et~al.} 2013,
  \nat, 493, 187

\bibitem[{{Middleton} {et~al.}(2018){Middleton}, {Walton}, {Alston}, {Dauser},
  {Eikenberry}, {Jiang}, {Fabian}, {Fuerst}, {Brightman}, {Marshall}, {Parker},
  {Pinto}, {Harrison}, {Bachetti}, {Altamirano}, {Bird}, {Perez},
  {Miller-Jones}, {Charles}, {Boggs}, {Christensen}, {Craig}, {Forster},
  {Grefenstette}, {Hailey}, {Madsen}, {Stern}, \& {Zhang}}]{Middleton1810}
{Middleton}, M.~J., {Walton}, D.~J., {Alston}, W., {et~al.} 2018, ArXiv
  e-prints, arXiv:1810.10518

\bibitem[{{Miller} {et~al.}(2013){Miller}, {Walton}, {King}, {Reynolds},
  {Fabian}, {Miller}, \& {Reis}}]{Miller1013}
{Miller}, J.~M., {Walton}, D.~J., {King}, A.~L., {et~al.} 2013, \apj, 776, L36

\bibitem[{{Mushtukov} {et~al.}(2017){Mushtukov}, {Suleimanov}, {Tsygankov}, \&
  {Ingram}}]{Mushtukov1705}
{Mushtukov}, A.~A., {Suleimanov}, V.~F., {Tsygankov}, S.~S., \& {Ingram}, A.
  2017, \mnras, 467, 1202

\bibitem[{{Mushtukov} {et~al.}(2015{\natexlab{a}}){Mushtukov}, {Suleimanov},
  {Tsygankov}, \& {Poutanen}}]{Mushtukov1512}
{Mushtukov}, A.~A., {Suleimanov}, V.~F., {Tsygankov}, S.~S., \& {Poutanen}, J.
  2015{\natexlab{a}}, \mnras, 454, 2539

\bibitem[{{Mushtukov} {et~al.}(2015{\natexlab{b}}){Mushtukov}, {Suleimanov},
  {Tsygankov}, \& {Poutanen}}]{Mushtukov1502}
---. 2015{\natexlab{b}}, \mnras, 447, 1847

\bibitem[{{Ohsuga} \& {Mineshige}(2011)}]{Ohsuga1107}
{Ohsuga}, K., \& {Mineshige}, S. 2011, \apj, 736, 2

\bibitem[{{Ohsuga} {et~al.}(2009){Ohsuga}, {Mineshige}, {Mori}, \&
  {Kato}}]{Ohsuga0906}
{Ohsuga}, K., {Mineshige}, S., {Mori}, M., \& {Kato}, Y. 2009, \pasj, 61, L7

\bibitem[{{Olausen} \& {Kaspi}(2014)}]{Olausen0514}
{Olausen}, S.~A., \& {Kaspi}, V.~M. 2014, The Astrophysical Journal Supplement
  Series, 212, 6

\bibitem[{{{\"O}zel} {et~al.}(2010){{\"O}zel}, {Psaltis}, {Narayan}, \&
  {McClintock}}]{Ozel1012}
{{\"O}zel}, F., {Psaltis}, D., {Narayan}, R., \& {McClintock}, J.~E. 2010,
  \apj, 725, 1918

\bibitem[{{Parfrey} \& {Tchekhovskoy}(2017)}]{Parfrey1712}
{Parfrey}, K., \& {Tchekhovskoy}, A. 2017, \apjl, 851, L34

\bibitem[{{Pintore} {et~al.}(2017){Pintore}, {Zampieri}, {Stella}, {Wolter},
  {Mereghetti}, \& {Israel}}]{Pintore1702}
{Pintore}, F., {Zampieri}, L., {Stella}, L., {et~al.} 2017, \apj, 836, 113

\bibitem[{{Pols} {et~al.}(1998){Pols}, {Schr{\"o}der}, {Hurley}, {Tout}, \&
  {Eggleton}}]{Pols9808}
{Pols}, O.~R., {Schr{\"o}der}, K.-P., {Hurley}, J.~R., {Tout}, C.~A., \&
  {Eggleton}, P.~P. 1998, \mnras, 298, 525

\bibitem[{{Popov}(2016)}]{Popov0116}
{Popov}, S.~B. 2016, Astronomical and Astrophysical Transactions, 29, 183

\bibitem[{{Poutanen} {et~al.}(2007){Poutanen}, {Lipunova}, {Fabrika},
  {Butkevich}, \& {Abolmasov}}]{Poutanen0705}
{Poutanen}, J., {Lipunova}, G., {Fabrika}, S., {Butkevich}, A.~G., \&
  {Abolmasov}, P. 2007, \mnras, 377, 1187

\bibitem[{{Rappaport} {et~al.}(2005){Rappaport}, {Podsiadlowski}, \&
  {Pfahl}}]{Rappaport0501}
{Rappaport}, S.~A., {Podsiadlowski}, P., \& {Pfahl}, E. 2005, \mnras, 356, 401

\bibitem[{{Sana} {et~al.}(2012){Sana}, {de Mink}, {de Koter}, {Langer},
  {Evans}, {Gieles}, {Gosset}, {Izzard}, {Le Bouquin}, \&
  {Schneider}}]{Sana1207}
{Sana}, H., {de Mink}, S.~E., {de Koter}, A., {et~al.} 2012, Science, 337, 444

\bibitem[{{S{\c a}dowski} {et~al.}(2014){S{\c a}dowski}, {Narayan}, {McKinney},
  \& {Tchekhovskoy}}]{Sadowski1403}
{S{\c a}dowski}, A., {Narayan}, R., {McKinney}, J.~C., \& {Tchekhovskoy}, A.
  2014, \mnras, 439, 503

\bibitem[{{S{\c a}dowski} {et~al.}(2015){S{\c a}dowski}, {Narayan},
  {Tchekhovskoy}, {Abarca}, {Zhu}, \& {McKinney}}]{Sadowski1502}
{S{\c a}dowski}, A., {Narayan}, R., {Tchekhovskoy}, A., {et~al.} 2015, \mnras,
  447, 49

\bibitem[{{Shakura} \& {Sunyaev}(1973)}]{Shakura73}
{Shakura}, N.~I., \& {Sunyaev}, R.~A. 1973, \aap, 24, 337

\bibitem[{{Shao} \& {Li}(2015)}]{Shao1504}
{Shao}, Y., \& {Li}, X.-D. 2015, \apj, 802, 131

\bibitem[{{Silva Aguirre} {et~al.}(2018){Silva Aguirre}, {Bojsen-Hansen},
  {Slumstrup}, {Casagrande}, {Kawata}, {Ciuc{\v a}}, {Handberg}, {Lund},
  {Mosumgaard}, {Huber}, {Johnson}, {Pinsonneault}, {Serenelli}, {Stello},
  {Tayar}, {Bird}, {Cassisi}, {Hon}, {Martig}, {Nissen}, {Rix},
  {Sch{\"o}nrich}, {Sahlholdt}, {Trick}, \& {Yu}}]{SilvaAguirre1804}
{Silva Aguirre}, V., {Bojsen-Hansen}, M., {Slumstrup}, D., {et~al.} 2018,
  \mnras, 475, 5487

\bibitem[{{Soria} {et~al.}(2014){Soria}, {Long}, {Blair}, {Godfrey}, {Kuntz},
  {Lenc}, {Stockdale}, \& {Winkler}}]{Soria1403}
{Soria}, R., {Long}, K.~S., {Blair}, W.~P., {et~al.} 2014, Science, 343, 1330

\bibitem[{{Swartz} {et~al.}(2004){Swartz}, {Ghosh}, {Tennant}, \&
  {Wu}}]{Swartz0410}
{Swartz}, D.~A., {Ghosh}, K.~K., {Tennant}, A.~F., \& {Wu}, K. 2004, \apjs,
  154, 519

\bibitem[{{Swartz} {et~al.}(2011){Swartz}, {Soria}, {Tennant}, \&
  {Yukita}}]{Swartz1111}
{Swartz}, D.~A., {Soria}, R., {Tennant}, A.~F., \& {Yukita}, M. 2011, \apj,
  741, 49

\bibitem[{{Swartz} {et~al.}(2009){Swartz}, {Tennant}, \& {Soria}}]{Swartz0909}
{Swartz}, D.~A., {Tennant}, A.~F., \& {Soria}, R. 2009, \apj, 703, 159

\bibitem[{{Tetarenko} {et~al.}(2016){Tetarenko}, {Sivakoff}, {Heinke}, \&
  {Gladstone}}]{Tetarenko1602}
{Tetarenko}, B.~E., {Sivakoff}, G.~R., {Heinke}, C.~O., \& {Gladstone}, J.~C.
  2016, The Astrophysical Journal Supplement Series, 222, 15

\bibitem[{{Tsygankov} {et~al.}(2017){Tsygankov}, {Doroshenko}, {Lutovinov},
  {Mushtukov}, \& {Poutanen}}]{Tsygankov1702}
{Tsygankov}, S.~S., {Doroshenko}, V., {Lutovinov}, A.~A., {Mushtukov}, A.~A.,
  \& {Poutanen}, J. 2017, ArXiv e-prints, arXiv:1702.00966

\bibitem[{{Vink}(2015)}]{Vink15}
{Vink}, J.~S. 2015, in Astrophysics and Space Science Library, Vol. 412, Very
  Massive Stars in the Local Universe, ed. J.~S. {Vink}, 77

\bibitem[{{Walton} {et~al.}(2011){Walton}, {Roberts}, {Mateos}, \&
  {Heard}}]{Walton1109}
{Walton}, D.~J., {Roberts}, T.~P., {Mateos}, S., \& {Heard}, V. 2011, \mnras,
  416, 1844

\bibitem[{{Walton} {et~al.}(2018){Walton}, {F{\"u}rst}, {Heida}, {Harrison},
  {Barret}, {Stern}, {Bachetti}, {Brightman}, {Fabian}, \&
  {Middleton}}]{Walton1804}
{Walton}, D.~J., {F{\"u}rst}, F., {Heida}, M., {et~al.} 2018, \apj, 856, 128

\bibitem[{{Wiktorowicz} {et~al.}(2014){Wiktorowicz}, {Belczynski}, \&
  {Maccarone}}]{Wiktorowicz1409}
{Wiktorowicz}, G., {Belczynski}, K., \& {Maccarone}, T. 2014, in Binary
  Systems, their Evolution and Environments, 37

\bibitem[{{Wiktorowicz} {et~al.}(2017){Wiktorowicz}, {Sobolewska}, {Lasota}, \&
  {Belczynski}}]{Wiktorowicz1709}
{Wiktorowicz}, G., {Sobolewska}, M., {Lasota}, J.-P., \& {Belczynski}, K. 2017,
  \apj, 846, 17

\bibitem[{{Wiktorowicz} {et~al.}(2015){Wiktorowicz}, {Sobolewska},
  {S\k{a}dowski}, \& {Belczynski}}]{Wiktorowicz1509}
{Wiktorowicz}, G., {Sobolewska}, M., {S\k{a}dowski}, A., \& {Belczynski}, K.
  2015, \apj, 810, 20

\bibitem[{{Wolter} {et~al.}(2018){Wolter}, {Fruscione}, \&
  {Mapelli}}]{Wolter1808}
{Wolter}, A., {Fruscione}, A., \& {Mapelli}, M. 2018, \apj, 863, 43

\bibitem[{{Xiang} {et~al.}(2018){Xiang}, {Shi}, {Liu}, {Yuan}, {Chen}, {Huang},
  {Wang}, {Wu}, {Tian}, {Huo}, {Zhang}, \& {Zhang}}]{Xiang1808}
{Xiang}, M., {Shi}, J., {Liu}, X., {et~al.} 2018, \apjs, 237, 33

\end{thebibliography}

\end{document}